\documentclass[Numbered,notitlepage,onecolumn,sn-aps]{utils/sn-jnl}

\usepackage{graphicx}%
\usepackage{multirow}%
\usepackage{amsmath,amssymb,amsfonts}%
\usepackage{amsthm}%
\usepackage{mathrsfs}%
\usepackage[title]{appendix}%
\usepackage{xcolor}%
\usepackage{textcomp}%
\usepackage{manyfoot}%
\usepackage{booktabs}%
\usepackage{algorithm}%
\usepackage{algorithmicx}%
\usepackage{algpseudocode}%
\usepackage{listings}%

\graphicspath{{Figures/}{figures/}}

\usepackage{tabularx}
\usepackage{subfig}
\usepackage{siunitx}
\usepackage{algpseudocode}
\usepackage{algorithm}
\usepackage{gensymb}
\newcommand{\etal}{\textit{et al.}}
\newcommand{\ddt}[1]{\frac{\partial {#1}}{\partial t}}
\renewcommand{\vec}{\boldsymbol}

\theoremstyle{thmstyleone}%
\theoremstyle{thmstyletwo}%

\theoremstyle{thmstylethree}%

\begin{document}

\title[Article Title]{GPU-accelerated Linear Algebra for Coupled Solvers in Industrial CFD Applications with OpenFOAM}


\author*[1]{\fnm{Stefano} \sur{Oliani}}\email{stefano.oliani.ext@leonardo.com} \email{stefanooliani95@gmail.com}
\author[1]{\fnm{Ettore} \sur{Fadiga}}
\author[1]{\fnm{Ivan} \sur{Spisso}}
\author[1]{\fnm{Luigi} \sur{Capone}}
\author[2]{\fnm{Federico} \sur{Piscaglia}}

\affil*[1]{\orgname{Leonardo S.p.a}, \orgaddress{\street{Via Pieragostini 80},
    \city{Genova}, \country{Italy}}}

\affil[2]{\orgdiv{Dept. of Aerospace Science and Technology},
  \orgname{Politecnico di Milano}, \orgaddress{\country{Italy}}}


\abstract{The present work describes the development of heterogeneous GPGPU implicit CFD coupled
  solvers, encompassing both density- and pressure-based approaches. In this
  setup, the assembled linear matrix is offloaded onto multiple GPUs using
  specialized external libraries to solve the linear problem
  efficiently. These coupled solvers are applied to two industrial test cases
  representing common scenarios: the NASA CRM in a transonic regime
  \cite{rivers2010} and the external aerodynamics study of the DriveAER car
  \cite{2014-01-0590}. Significant performance enhancements are evident when
  compared to their CPU counterparts. Specifically, the NASA CRM case achieves
  an overall speedup of more than 4x, while the DriveAER test case
  demonstrates improved stability and reduced computational time compared to
  segregated solvers. All calculations were carried out utilizing the
  GPU-based partition of the \texttt{davinci-1} supercomputer at the Leonardo
  Labs, featuring 82 GPU-accelerated nodes.}

\keywords{GPU Acceleration, implicit density-based coupled solver, implicit
  pressure-based coupled solver, OpenFOAM, AmgX}

\maketitle

\section{Introduction}

The justification for utilizing GPUs in CFD simulations lies in their highly
efficient massively parallel technology and impressive performance per watt
\cite{6270749,Price2016,KRZYWANIAK2023396}. Given the distinctive
architectural variances between GPUs and CPUs, the development of specialized
algorithms becomes imperative to fully harness the potential performance of
GPU hardware \cite{Ansys, starCCM, naumov2015, piscaglia2023, WILLIAMS2016,
  ZHANG2023}. Given the relatively recent emergence of this field, there
remains significant scope for enhancing the acceleration of these solutions.

There are primarily two approaches to accelerate CFD simulations. The first
approach entails a comprehensive overhaul of the CFD code, specifically
optimized for exclusive execution on Graphics Processing Units
(GPUs). Although this method often yields significant speed improvements, it
comes with trade-offs. It typically demands a considerably longer development
period and is highly tailored to address a specific problem, referred to as a
``full port''. The alternative approach involves adopting a partial-offloading
strategy, where only the computationally intensive portions of the code are
transferred to the GPU. There are already many examples in which fast GPU
simulations have been demonstrated for unstructured finite-volume CFD software
using the full or partial offloading strategy \cite{piscaglia2023,
  WILLIAMS2016, jaiswal2016, ZHANG2023}.

In CFD software packages, steady-state simulations for external aerodynamics
are typically carried out using either pressure-based or density-based
solvers. The choice between these solvers depends on the flow speed. While
implicit density-based solvers present a coupled structure by nature,
pressure-based solvers can be divided into SIMPLE-based and coupled
methods. The SIMPLE operation follows a segregated approach, addressing linear
systems for individual physical variables like pressure and velocity
components, one at a time. While this approach has limited memory
requirements, it often results in slow convergence, requiring a substantial
number of iterations \cite{Peric}. When implementing a partial port strategy
for SIMPLE-based solvers, a significant portion of the overall solution time
is spent transferring data between the host (CPU) and the GPU for each linear
solver. Consequently, the overall performance gains from this strategy tend to
be limited to around a 2x speedup \cite{piscaglia2023}. To overcome these
limitations, the exploration of coupled solvers \cite{mangani2014,
  mangani2016, DARWISH2018} emerges as a promising avenue for implementing the
partial port strategy on GPUs. Compared to segregated solvers, coupled solvers
offer notable advantages, including linear scalability of CPU time with cell
count and the capacity to achieve more rapid convergence of residuals,
particularly for steady-state flows. Implicit coupled solvers typically
combine robustness, stability and a reduced number of nonlinear iterations
required for the convergence, especially for steady-state problems. These
features result in a significantly improved overall convergence of the
solver. This statement is counter-balanced by the increased size of a linear
system of algebraic equations that needs to be solved in the coupled solution,
as well as the increase in computer storage and computational effort per block
solution. Current advancements in computer technology have led to the
availability of powerful machines and a strong push towards achieving exascale
performance in numerical simulation codes. These developments have
significantly altered the trade-off between memory requirements and simulation
speed, particularly when dealing with computational meshes that consist of
hundreds of millions of computational points. 

Historically, the practical application of coupled solvers on GPUs in
large-scale industrial settings faced a significant hurdle due to the
traditional constraints on GPU hardware's memory capacity. The motivation for
the current development work lies in the focus on the attention to modern GPU
cards, which are characterized by enhanced memory capacity and adeptness in
handling the computational demands of industrial workflows
\cite{nvidiaH100}. This ongoing effort is therefore centered on the adaptation
of existing CFD workflows to fully exploit the capabilities offered by
contemporary GPU technology. Open-source tools are proving to be valuable
assets for industrial CFD applications.  In this context, OpenFOAM
\cite{weller1998} is the most widely employed open-source tool for CFD
simulations in Academia and Industry \cite{audiOF}. In the author' knowledge,
despite several effort have been made of either density- and pressure-based
implicit coupled solvers in OpenFOAM \cite{OLIANI2023, UROIC2021,
  mangani2014,DARWISH2009}, implicit coupled solvers are not available in the
official distributions of the software \cite{OpenFOAM-com,OpenFOAM-dev}. While
significant efforts have been made to accelerate linear algebra in
SIMPLE-based solvers \cite{piscaglia2023} and to accelerate combustion
calculations \cite{GHIOLDI2023}, there are currently no investigations of
performance obtained through partial offload strategy on GPUs of coupled
solvers in the OpenFOAM framework.

We present a study on the performance of implicit coupled linear algebra GPU
offloading for the \texttt{ICSFoam} library \cite{OLIANI2023}. The study
encompasses results obtained from both an implicit compressible density-based
solver and an incompressible pressure-based coupled solver. These solvers are
applied to two industrial test cases that represent typical
scenarios. Specifically, the implicit compressible density-based coupled
solver is applied to the NASA CRM in a transonic regime \cite{rivers2010},
while the pressure-based implicit coupled solver is employed for the external
aerodynamics of the DriveAER car \cite{2014-01-0590}. The goal of this work is
to assess the acceleration of a GPGPU CFD solver based on the OpenFOAM
framework. In this approach, the algebraic solution of the assembled coupled
matrix from the discretization of the flow transport problem is offloaded onto
multiple GPUs. This strategy is expected to significantly enhance the
computational performance of external aerodynamics simulations across various
flow speeds.

The paper is structured as follows: in Sec. \ref{sec:met}, we provide a brief
overview of the methodology used to assemble coupled matrices for the pressure
and density-based algorithms employed in this study. Additionally, we outline
the strategy adopted for GPU offloading of linear algebra using the NVIDIA
\texttt{AmgX} library \cite{naumov2015}. Sec. \ref{sec:hardware} provides a
summary of the hardware and techniques utilized for the performance
assessment. In Sec. \ref{sec:res}, we present the results for two test cases with industrial relevance. The simulations are initially validated by comparing them to
experimental data and other computational fluid dynamics (CFD)
solutions. Subsequently, the performance benefits from the presented
implementation are discussed. Finally, conclusions from the study are drawn in
Sec. \ref{sec:concl}.

\section{Methodology} \label{sec:met}

An in-house version of the \textit{ICSFoam} library \cite{OLIANI2023} was used
for this work. The library was developed starting from the \textit{HiSA}
solver \cite{heyns2014} as a generalization of OpenFOAM for arbitrary systems
of implicitly coupled partial differential equations. A detailed description
can be found in \cite{OLIANI2023}. The library was initially designed for
compressible turbomachinery flows \cite{oliani2022HBM}, but was also employed
for multiphysics aeroelastic simulations of transonic flutter
\cite{fadiga2023}. In this section, a broad overview of the coupled
pressure-based and density-based solvers used in this work is given before
describing the linear algebra acceleration through \textit{AmgX}.

\begin{figure}[H]
  \centering \includegraphics[width=.7\textwidth]{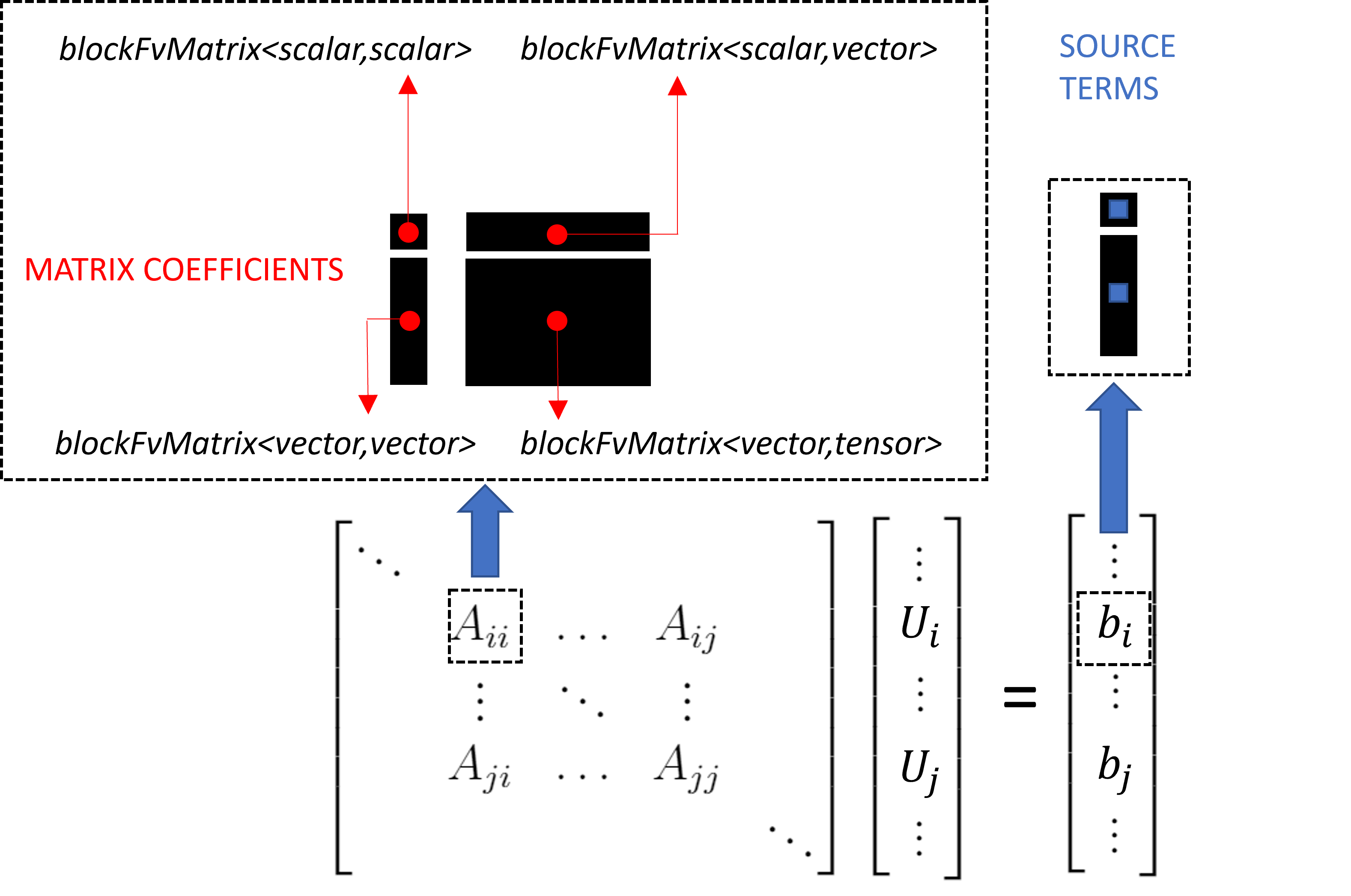}
  \caption{Sub-blocks and source term structure for a system of coupled
    equations with one scalar and one vector variables.}
  \label{fig:subblocks}
\end{figure}

In general, a linearized and discretized coupled system of equations is
written as:

\begin{equation}
  \label{eq:coupl_system}
  A_{ii} \vec{U_i} + \sum_N A_{ij} \vec{U_j} = \vec{b_i}
\end{equation}

for a generic cell $i$ of the mesh and its neighbours $N$. The coefficients,
or blocks, can be seen as small sub-matrices of dimension $n \times n$, where
$n$ is the number of equations. In turn, each block is split into sublocks
based on the coupling coefficients in terms of scalar and vector
variables. This allows a very intuitive treatment of coupled systems and can
be easily generalized to any number of variables. As an example,
Fig. \ref{fig:subblocks} shows this splitting for the incompressible
Navier-Stokes equations, where one scalar and one vector equations are
present.

For the solution of linear systems of equations in the CPU version of the
code, the GMRES Krylov subspace solver \cite{saad1986} combined with the Lower
Upper Symmetric Gauss-Seidel \cite{jameson1981} preconditioner are used. This
combination has proven a suitable approach in terms of efficiency and
stability in density-based solvers for compressible Navier-Stokes equations.

\subsection{Density-based solver}

The three-dimensional compressible Reynolds-averaged Navier-Stokes (RANS)
equations in conservation form can be written as:

\begin{equation}\label{navStokesEq}
  \int_V \ddt{\vec{Q}} dV +\int_{\partial V} (\vec{F_c - F_v}) dS = 0
\end{equation}

where

\begin{equation}
  \vec{Q} =
  \begin{bmatrix}
    \rho \\ \rho \vec{u} \\ \rho E
  \end{bmatrix}
  \!  , \vec{F_c} =
  \begin{bmatrix}
    \rho \vec{u} \cdot \vec{n} \\ (\rho \vec{u} \otimes \vec{u}) \cdot \vec{n}
    + p \vec{n}\\ \rho \vec{u}H \cdot \vec{n}
  \end{bmatrix}
  \!  , \vec{F_v} =
  \begin{bmatrix}
    0 \\ \vec{\tau} \cdot \vec{n}\\ (\vec{\tau} \cdot \vec{u} + \vec{q}) \cdot
    \vec{n}
  \end{bmatrix}
\end{equation}

and $\vec{n}$ is the face normal vector, $\rho$ is the density, $\vec{u}$ is
the velocity, $E$ is the total internal energy, $H$ is the total enthalpy, $p$
is the static pressure, $\vec{\tau}$ is the viscous stress tensor and
$\vec{q}$ is the heat flux vector. In the case of steady-state equations, the
time derivative is replaced with a pseudo-time term $\tau$ in the
equations. Using the finite volume method for the spatial discretization of
Eqs. \eqref{navStokesEq}, we get a set of semi-discretized equations:

\begin{equation}
  \label{semidiscrEq}
  V \frac{\partial \vec{Q}}{\partial \tau} = \vec{R(Q)}
\end{equation}

When an implicit method is employed to march the equations in pseudo-time to
the iteration $n+1$, the residual is linearized about iteration $n$ as

\begin{equation}
  \vec{R(Q^{n+1})} \approx \vec{R(Q^n)} + \frac{\partial
    \vec{R(Q)}}{\partial \vec{Q}} \biggr|_{\vec{Q=Q^n}} \Delta \vec{Q^n},
  \qquad \Delta \vec{Q^n} = \vec{Q^{n+1}} - \vec{Q^n}
\end{equation}

At each iteration the following linear system is solved to compute the
solution increment $\Delta \vec{Q^n}$:

\begin{equation}
  \label{eq:discretBackward}
  \biggl[\frac{V}{\Delta \tau} \mathbf{I} - \frac{\partial
      \vec{R(Q^n)}}{\partial \vec{Q^n}}\biggr] \Delta \vec{Q^n} =
  \vec{R(Q^n)}
\end{equation}

Standard Newton iterations involving exact linearizations of the numerical
flux $\vec{R(Q)}$ would give the best convergence rate for the nonlinear
equations \eqref{semidiscrEq}, but are very expensive to compute, if not
impossible to evaluate. Therefore, approximate Jacobians are usually adopted
in the linearization of the implicit part. In this work we adopt the
approximate Jacobian formulation of Luo \etal \cite{luo1998}, so that the
contribution of each internal face to block coefficients can be expressed as:

\begin{gather}
  \label{eq:approxJac1}
  A_{ii} = A_{ii} + \frac{\partial \vec{R_{ij}(n_{ij})}}{\partial \vec{Q_i}}
  = A_{ii} + \frac{1}{2} S_f (\vec{J(Q_i, n_{ij})} +
  |\lambda_{ij}|\mathbf{I}) \\
  \label{eq:approxJac2}
  A_{ij} = A_{ij} + \frac{\partial \vec{R_{ij}(n_{ij})}}{\partial \vec{Q_j}}
  = A_{ij} + \frac{1}{2} S_f (\vec{J(Q_j, n_{ij})} -
  |\lambda_{ij}|\mathbf{I})
\end{gather}

where $S_f$ is the face area, $\vec{J}$ is the convective flux Jacobian and
$\lambda_{ij}$ is the sum of the spectral radii of the Roe and viscous flux
matrices (see \cite{blazek2015} for more details). The dependence on the
surface outward-normal vector $\vec{n_{ij}}$ pointing from cell $i$ to cell
$j$ has been highlighted since caution must be taken according to the sign of
the flux. Indeed one has that: $\vec{R_{ij}(n_{ij})} = -
\vec{R_{ji}(n_{ji})}$.

The MUSCL reconstruction-evolution approach is used to calculate the numerical
flux residual term $\vec{R(Q_i)}$. The first step involves primitive variables
interpolation from cell centroids to face centers. TVD slope-limiters are used
in the vicinity of shocks to avoid the appearence of spurious oscillations
that may lead to stability issues. After the reconstruction, left and right
states are used to compute the numerical flux through approximate Riemann
solvers at each cell interface. Roe, HLLC and AUSM+Up approximate Riemann
solvers are available in the ICSFoam library. Finally, integrating the fluxes
over all faces of the cell (cfr. second term of Eq.\ref{navStokesEq}) we
obtain the numerical flux residual $\vec{R(Q_i)}$.

\subsection{Pressure-velocity coupling for pressure-based solver}
Originally, pressure-based algorithms were developed for incompressible flows
in their segregated SIMPLE formulation \cite{patankar1980} and then extended
to the compressible regime \cite{ISSA1986}. They were originally developed in
the 1970s and 1980s, when parallelism was limited, and computational resources
were scarce. Consequently, numerical algorithms based on a segregated solution
procedure gained popularity. In segregated algorithms, the pressure–velocity
system are assembled by decoupling the momentum and pressure equations by
treating the other unknown explicitly, using the value from the previous
iteration; the equation set is solved sequentially, component by component and
equation coupling is achieved through Picard iterations. Optimal segregated
algorithms hold only a single matrix in memory and re-use its storage
space. Such treatment of the pressure–velocity system is unnatural since the
connection between the two variables is linear and can be resolved
simultaneously in a single linear system. To achieve this goal, we want first
to assemble a single linear system across the complete set of governing
equations to simultaneously: a) couple p and u fields without simplification;
b) remove the need for “inner” Picard iterations for p-u coupling and c)
remove the need for under-relaxation.

Extensive literature is available on the derivation of coupled pressure-based
solvers equations and the relative implementations. Incompressible and
compressible formulations can be found for OpenFOAM as well \cite{mangani2014,
  mangani2016}, even if no coupled solvers are currently available in the
official release. Here, for the sake of simplicity, we will limit ourselves to
the incompressible steady-state Navier-Stokes equations. Compressible
pressure-based solvers are usually obtained by solving a segregated energy
equation.

According to Eq.\eqref{eq:coupl_system}, the blocks of the system can be
written explicitly as:

\begin{equation}
  \label{eq:prb_coeff}
  A_{ii} =
  \begin{bmatrix}
    \vec{a}_{ii}^{uu} & \vec{a}_{ii}^{up} \\ \vec{a}_{ii}^{pu} &
    a_{ii}^{pp}
  \end{bmatrix}
  \qquad A_{ij} =
  \begin{bmatrix}
    \vec{a}_{ij}^{uu} & \vec{a}_{ij}^{up} \\ \vec{a}_{ij}^{pu} &
    a_{ij}^{pp}
  \end{bmatrix}
\end{equation}

Where the sub-blocks splitting previously described is evident and:
$\vec{a}^{uu}$ is a 3x3 tensor representing the diagonal term of the momentum
equation, $\vec{a}^{up}$ and $\vec{a}^{pu}$ are vectors representing the
off-diagonal coupling terms, and $a^{pp}$ is a scalar representing the
diagonal term of the pressure equation. In this section we derive formulas for
these terms in the case of pressure-velocity coupling.

Following the derivation of Mangani \etal \cite{mangani2014}, the steady
discretized momentum and continuity equations can be written as

\begin{gather}
  \sum_N \vec{S_f} \Bigl((\vec{u}_f \cdot \vec{n}) \vec{u}_f + p_f -
  \nu_{eff} \nabla \vec{u}_f\Bigr) = 0 \\ \sum_N \vec{S_f} \cdot \vec{u}_f =
  0
\end{gather}

Where the subscript $f$ indicates values at cell faces and $\nu_{eff}$
represents the sum of laminar and turbulent viscosity. The convective term is
linearized using $\vec{u}_f \cdot \vec{n}$ evaluated using previous iteration
values. Here, due to the assumption of constant density, the pressure $p$ is
assumed to be normalized by the density $\rho$ and therefore has dimension
\si{\meter^2/\second^2}. The pressure equation is obtained using the
discretized form of the continuity equation combined with the Rhie-Chow
interpolation \cite{rhie1983}:

\begin{equation}
  \sum_N \vec{S_f} \cdot \Bigl[\overline{\vec{u}_f} - \overline{\vec{D}_f}
    \cdot \bigl( \nabla p_f - \overline{\nabla p_f} \bigr) \Bigr] = 0
\end{equation}

Where the overbar denotes linear interpolation from cell centroids to face
centers and the operator $\vec{D} = V(\vec{a}_{ii}^{uu})^{-1}$, with $V$ being
the volume of the cell. This is needed to avoid checkerboard oscillations on
collocated grids and results from the discretized momentum equation
reconstructed at cell faces. We use $||\vec{u}_f||$ to indicate a generic
first or second-order upwind discretization of the convective flux in the
momentum equation. The surface-normal velocity gradient in the diffusive part
of the momentum equation and the implicit pressure gradient in the pressure
equation are evaluated using classical central-differencing with an explicit
non-orthogonal correction for unstructrured grids. For a detailed derivation
of these terms, see \cite{DARWISH2009}. With the above notation, the
contribution of each internal face to the block coefficients is:

\begin{gather*}
  \vec{a}_{ij}^{uu} = \Bigl(-||-\vec{u}_f|| - \nu_{eff} \frac{\vec{S}_f
    \cdot \vec{n}}{\vec{n} \cdot \vec{d_{ij}}} \Bigr)\mathbf{I} \qquad
  \vec{a}_{ii}^{uu} = \Bigl(||\vec{u}_f|| + \nu_{eff} \frac{\vec{S}_f \cdot
    \vec{n}}{\vec{n} \cdot \vec{d_{ij}}} \Bigr)\mathbf{I}
  \\ \vec{a}_{ij}^{up} = \vec{a}_{ij}^{pu} = (1-f_x) \vec{S_f} \qquad
  \vec{a}_{ii}^{up} = \vec{a}_{ii}^{pu} = f_x \vec{S_f} \\ a_{ij}^{pp} =
  \frac{(\overline{\vec{D}_f} \vec{S_f}) \cdot \vec{n}}{\vec{n} \cdot
    \vec{d_{ij}}} \qquad a_{ii}^{pp} = -a_{ij}^{pp}
\end{gather*}

where $f_x$ represents the linear interpolation weights and $\vec{d_{ij}}$ is
the distance vector between centroids of cells $i$ and $j$.

\subsection{AmgxCoupled4Foam}

Even if performances of native GPU solvers can't be achieved, offloading only
the linear algebra offers several advantages. First of all, a much smaller
effort is needed in term of implementation to create the necessary coupling
layer between the CFD solver and the external linear algebra library. Second,
the higher degree of flexibility and the possibility to test a wide variety of
linear solvers. External linear algebra libraries offer a comprehensive set of
linear solvers with competitive implementations, which are not usually
available in proprietary CFD solvers. Third, the solution of the linear system
is typically one of the most time-consuming part of the iteration, therefore
accelerating this portion of the code alone allows to obtain relevant
speedups.

As of now, one of the few concrete example of cutting-edge technology in the
field of linear algebra solvers for GPUs is provided by NVIDIA through their
open-source library, \textit{AmgX} \cite{naumov2015}. The \textit{AmgX}
library was chosen because it provides a comprehensive set of tools for
solving large sparse system of equations arising from the discretization of
differential equations on unstructured meshes. It features aggregation based
algebraic multigrid methods for block coupled systems, along with several
smoothers and preconditioners such as Gauss-Seidel, and incomplete-LU
factorization. These can be combined with a variety of preconditioned Krylov
subspace solvers are available, including GMRES and bi-conjugate
gradient. Parallelism is achieved thanks to graph matching techniques and
graph coloring algorithms. Moreover, the support already exists for the
coupling with OpenFOAM segregated solvers \cite{martineau2020}, and has
recently been tested for external aerodynamics and reactive flow simulations
\cite{piscaglia2023,GHIOLDI2023}. Therefore, the coupling layer only needed to
be adapted for coupled systems of equations and the underlying matrix
structure of the ICSFoam library. Moreover, the library has already been
tested successfully on the commercial solver ANSYS Fluent, leading to
significant performance improvement for the solution of p-U coupling in
pressure-based solvers. Indeed, finite-volume discretizations usually involve
a compact stencil where only direct face-neighbours of the cell are involved,
generating a very sparse matrix structure. Despite this fact, the matrix
blocks are usually dense. This makes possible to leverage coalesced memory
access on the GPU and obtain a more favorable arithmetic intensity compared to
segregated solvers, achieving higher speedups of the linear system
solution. In addition, since the system sparsity is not modified by the
coupled or segregated nature of the solution, LDU addressing as well as row
pointers and column indices for the CSR format are unchanged - the only
difference being that now they refer to block entries instead of scalar
coefficients. For scientific and computational applications, the CSR format
can help to save memory and improve computational efficiency in operations
involving large sparse matrices. CSR allows to access the matrix elements by
saving only the indices of the first element of each row. This reduction in
the number of indices translates directly to a reduction of the amount of data
that have to be transferred from the main memory to the processors during
Sparse Matrix Vector Multiplication (SpMVM) operations. The CSR matrix format
is compatible with NVIDIA AmgX.

The library presented in this paper, named \textit{amgxCoupled4Foam}, is
linked dynamically at runtime when the simulation is started and allows to
handle arbitrary coupled systems of equations through NVIDIA AmgX library on
GPU accelerators. The v2112 ESI release of OpenFOAM is used in combination
with CUDA Toolkit 11.4. The code structure follows best practice OpenFOAM
coding guidelines, using a virtual constructor to select at runtime the
desired linear solver. This allows, for example, to switch seamlessly between
CPU legacy and AmgX GPU linear solvers without stopping the simulation.

The first aspect that must be taken into account is the flexibility given by
the heterogeneous CPU/GPU paradigm. Consolidation of matrix partitions can be
used to run simulations with an arbitrary number of GPUs and MPI ranks,
provided the second is larger that the first. This is important because the
part of the code that runs on the host can still benefit from increasing the
number of CPU cores, without the necessity to increase the number of GPU
devices as well. Indeed, several matrix partitions generated by different MPI
ranks are consolidated on a root rank that sends compute requests to a single
GPU device. This can be easily achieved by offsetting properly the row and
column indices of each matrix partition to place the coefficients in the right
position inside the consolidated matrix. This allows to fully benefit from
increasing the number of CPU cores and GPU devices independently, balancing
them to the user's discretion. This type of flexibility of the heterogeneous
approach finds no counterpart in the native GPU implementations. In the latter
scenario, there is typically one MPI rank for each GPU device, so that the
code must be completely revamped to port every single kernel on the
accelerator. If this is not the case, even small code routines that are not
accelerated can become significant bottlenecks in the code.

\begin{figure}
  \centering \includegraphics[width=.9\textwidth]{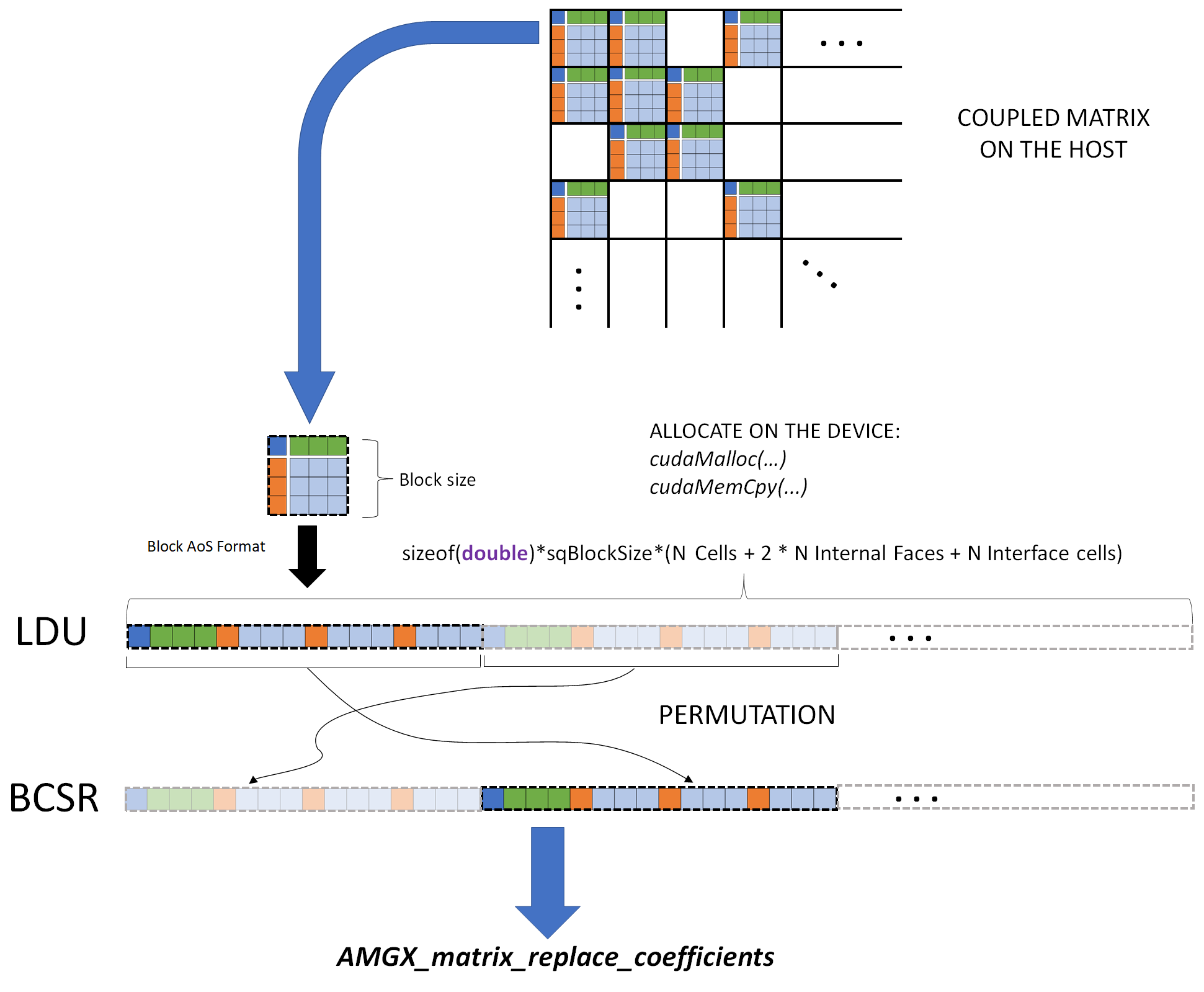}
  \caption{Schematic procedure of coupled matrix values conversion into an array
    in block AoS format. Permutation of the blocks is needed to switch from
    OpenFOAM native LDU storage to block CSR format needed in AmgX. }
  \label{fig:block_amgx}
\end{figure}

Algorithm \ref{alg:amgxFoam} describes the steps needed to correctly transfer
matrix coefficients to Amgx. Attention needs to be taken to fully exploit the
block-coupled matrix structure in order to obtain better performances in the
solution of the linear system. AmgX needs CSR block AoS format to carry out
the solution of the linear system. This procedure is described schematically
in Fig. \ref{fig:block_amgx}. In the upper part of the figure, a small portion
of the whole matrix is sketched, highlighting the division into
sub-blocks. Incompressible Navier-Stokes equations are used as an example
($4\times4$ blocks). Looping over each block of the matrix, the coefficients
are linearized into an array in row-major format. Blocks are then concatenated
to each other to obtain the array of coefficients in block AoS format. This
array is then allocated and copied on the device, and a permutation CUDA
kernel based on row and column indices is applied to convert the array to CSR
format. A similar procedure, is also applied to convert the solution and the
source term of the system in block AoS format, obtaining scalar array
pointers. Permutation is not needed in this case since the pattern is implied.

\begin{algorithm}
  \caption{Procedure for coupled system solution offloading on AmgX}
  \label{alg:amgxFoam}
  \begin{algorithmic}[1]
    \State Convert source term and solution array from PtrLists of scalar and
    vector fields to block AoS format.  \If{First iteration} \State Initialize
    AmgX with linear solver options and pass MPI communicators data to AmgX \State
    Map available devices on MPI ranks used by the application for consolidation
    \EndIf

    \State Convert coupled matrix coefficients into block Aos format
    (Fig. \ref{fig:block_amgx}) \State Exchange values with neighbour processor
    patches through MPI communications.

    \If{(First iteration or mesh topology has changed)} \State Setup rank-local
    and global number of rows so that AmgX can deal with matrix partitions \State
    Setup AmgX operations by uploading distributed row offsets, column indices and
    matrix values.  \Else \State Replace the old matrix coefficients with new ones
    \EndIf

    \State Permute matrix coefficients to obtain block CSR from LDU format.

    \State Upload solution and source term arrays on the device taking into
    account row offsets due to consolidation \State Call AmgX solver to solve the
    linear system \State Download solution array in block AoS format on the
    host. Each rank must read its portion of the consolidated solution.  \State
    Get linear solver number of iterations and residuals on root rank and output
    them \State Retrieve solution array as PtrLists of scalar and vector fields
    from block AoS format

    \If{Final iteration} \State Free memory on the device and destroy AmgX objects
    \EndIf

  \end{algorithmic}
\end{algorithm}

\begin{figure}
  \centering \includegraphics[width=\textwidth]{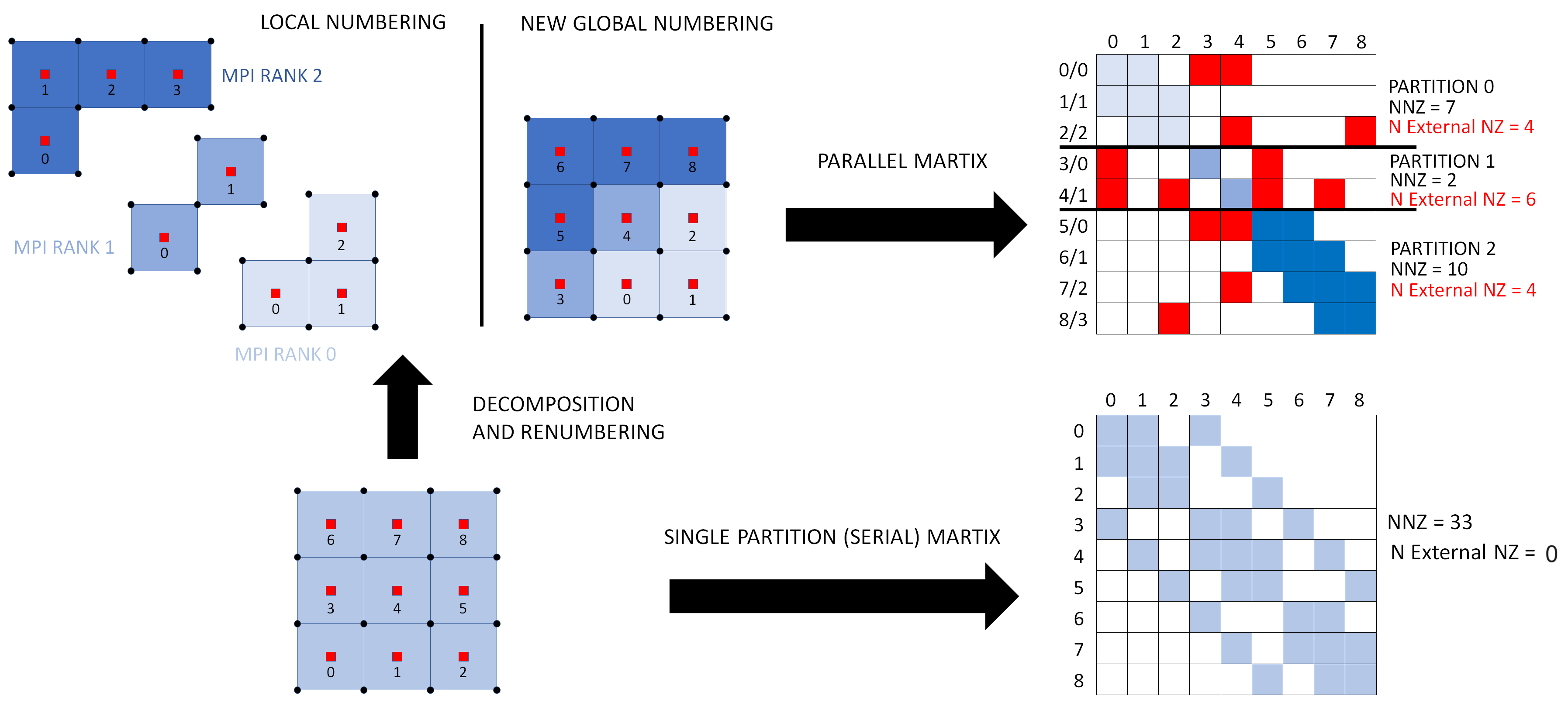}
  \caption{Procedure for matrix partitioning and coupled processor patches
    coefficients insertion in distributed parallel matrix}
  \label{fig:matrix_part}
\end{figure}

An important remark concerns the exchange of boundary values on coupled
patches, especially processor interfaces. Fig. \ref{fig:matrix_part}
schematically shows the procedure for a simple domain with nine cells
partitioned into three MPI ranks. In the lower part of the figure, the matrix
sparsity pattern is shown for the serial matrix. For distributed matrices,
arbitrary partitioning (obtained e.g. with SCOTCH or METIS) are dealt with by
renumbering the mesh after decomposition. Halo layers are transferred from
neighbour processors and included into the present matrix partition in CSR
format using the local row number and global column index (in red in the
figure). These coefficients are stored into arrays of size $NExternalNZ$ equal
to the number of processor faces shared with neighbour partitions. For serial
runs, this array is obviously empty, in absence of other kinds of coupled
patches (e.g. cyclics). Please notice that, due to renumbering, the matrix
sparsity pattern can change. This approach, however, makes the management of
the distributed matrix much easier since each processor owns complete
consecutive rows, so that there is no need to build and pass a partition
vector to AmgX. Instead, a trivial partitioning is implied where rows are
consecutive in each rank and this is true independently of how the mesh is
decomposed. The same approach is used for periodic conformal interfaces
(\textit{cyclics} in OpenFOAM). In such case, eventual rotations between the
coupled patches must be taken into account by multiplying the matrix
coefficient by the corresponding transformation matrix.

\section{Hardware and Applications}
\label{sec:hardware}
This section contains a general description of the testing procedure that has
been applied to the libraries developed in the scope of this paper. In
particular, the computing hardware is presented, together with a set of
application cases that have been selected to evaluate the performance, the
accuracy and the stability of the solvers.

The hardware selected for the computations is a set of GPU-accelerated nodes
of the davinci-1 supercomputer \cite{davinci-1}. This HPC system, recently
installed at one of the Leonardo facilities in Genova, is equipped with 82
GPU-accelerated nodes characterized by:
\begin{itemize}
\item 2x24 cores AMD EPYC Rome 7402 Processor
\item 512 GB total DRAM memory
\item 4 x A100-SXM4-40GB NVIDIA GPUs
\item 9.746 TFLOPS in FP64 (double) performance
\item PCIe 4.0 x8 bus connection
\item InfiniBand switches for nodes interconnection
\end{itemize}

More precisely, the authors have employed up to 18 nodes to test the
performance and the scalability of both the pressure-based and the
density-based solver. The availability of a considerable amount of
computational resources has allowed the selection of application cases
characterized by a level of complexity typical of industrial applications.

In section \ref{sec:res}, the results of two open test cases will be presented
to assess the correct implementation and the performance of the coupled
solvers with GPU-accelerated linear algebra (GPU-LA from now on). The two
cases are used to benchmark the density and pressure-based solvers, and
present globally-accepted, widely adopted industrial geometries in the
aerospace and automotive fields, respectively. These have been chosen from two
widely known CFD workshops, so that comparison with available experimental
data as well as results obtained with other state-of-art CFD software is
possible.

Performance measurements are performed in a consistent manner for the two
selected cases. First, simulations are run until convergence, monitoring
residual values as well as aerodynamic coefficients. Then, the statistics are
recorded for the following 500 iterations. Profiling is done by simply
annotating the code outside the main computational kernels. For the present
work, only the coupled system is offloaded on the device, while segregated
turbulence equations are kept on the CPU. Future tests will investigate
performance gaining attainable by offloading also turbulence equations.

It must be emphasized that the CPU version of the ICSFoam library is mainly a
research code and is non-optimized for performances on many-core
architectures. Therefore, it does not feature the most competitive
implementation of iterative linear solvers. However, previous tests
\cite{OLIANI2023} showed that the CPU version of the code, though in general
slower than other CFD codes, leads to acceptable computational times that are
of the same order of magnitude of commercial codes. Put another way, the
information on the speedup obtained with the GPU-LA version can be considered
as fairly general and provides valuable insights into the attainable overall
acceleration due to linear algebra offloading.

\section{Results} \label{sec:res}

\subsection{Test Case 1: NASA CRM in transonic regime}

The first test case presented regards the simulation of the NASA Common
Research Model (CRM) \cite{vassberg2008} in the transonic flight
configuration, from the AIAA Drag Prediction Workshop (DPW) 6. The CRM
represents a modern commercial transport airplane and was designed as a full
configuration with a low wing, body, horizontal tail, and engine nacelles
mounted below the wing. A summary of results obtained by the participants can
be found in Tinoco \etal \cite{tinoco2018}. The full-scale geometry CAD is
available on the workshop website \cite{dpw6}, together with a number of
overset, structured and unstructured meshes generated by the participants and
by the committee. The wing-body-nacelle-pylon configuration was selected for
the present study, also considering experimentally measured static aeroelastic
deflections for an angle of attack of $2.75$ degrees. This has been shown to
dramatically improve predicted wing pressure distribution at transonic flight
conditions \cite{tinoco2018}. Half-span of the full-scale model is simulated,
by setting a symmetry condition on the corresponding plane. The domain is
hexahedrical with farfield surfaces placed 100 reference chords away from the
aircraft. The simulated conditions are reported in Table \ref{tab:BC_CRM}, and
aim at finding the predicted drag on the aircraft at a fixed lift coefficient
$C_L=0.5$. Two meshes generated by participants to the workshop were employed
in this work: a coarse mesh of 45 million elements for scalability tests and
comparisons between different linear solvers, and a fine mesh of about $260$
million elements for comparison with experimental data and other solvers. Both
meshes are unstructured and composed of mixed elements (hexahedra and
tetrahedra with prismatic layers). Freestream conditions are imposed at far
field boundaries, while a no-slip adiabatic condition is set at the
walls. Simulations are run fully turbulent in “free air” configuration,
without wind tunnel walls or support system. The baseline one-equation
Spalart-Allmaras turbulence model \cite{spalart1992} without rotation
correction is employed.

\begin{table} [h]
  \label{tab:CRM_BC}
  \centering
  \caption{Simulated conditions for transonic CRM test case. \label{tab:BC_CRM}}
  \begin{tabularx}{0.4\textwidth}{lll}
    \textbf{Quantity} & \textbf{Value}\\ \midrule Mach Number & 0.85 \\ $T_{ref}$
    & \SI{300}{\kelvin} \\ $p_{ref}$ & \SI{101325}{\pascal} \\ $Re$ & $5e6$
    \\ $C_L$ & 0.5 \\ Static Aeroelatic Deformation & $2.75 \degree$
    \\ $\tilde{\nu}/\nu_{\infty}$ & 3 \\ \bottomrule
  \end{tabularx}
\end{table}

\begin{figure}
  \includegraphics[width=.9\textwidth]{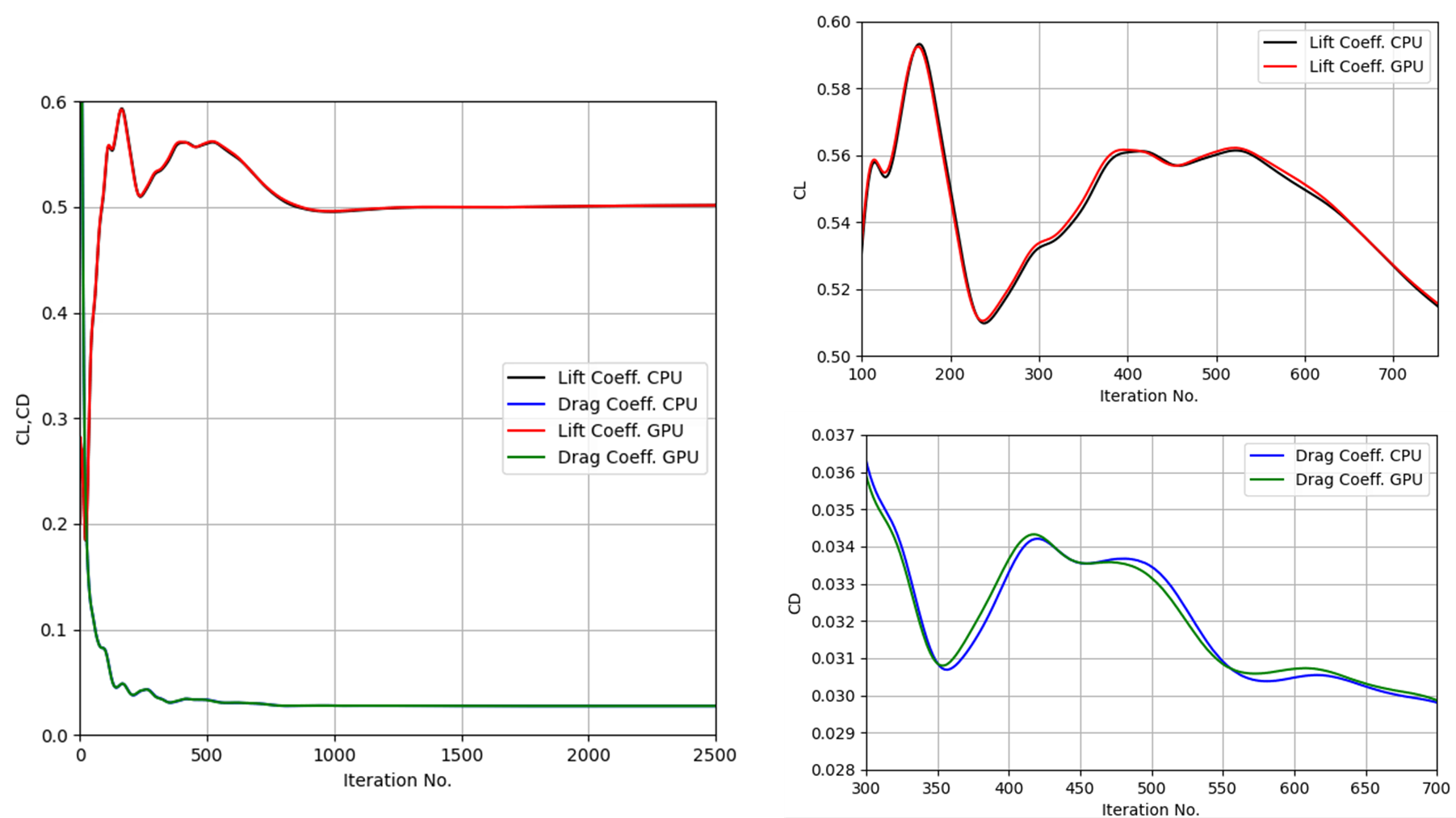}
  \caption{Aerodynamic coefficients convergence history for the first 2500
    iterations. CPU and GPU-offloaded linear algebra are represented. On the top
    and bottom right, closeup of lift and drag coefficient are shown,
    respectively.}
  \label{fig:crm_conv}
\end{figure}

First of all, preliminary tests are conducted on the coarse mesh to verify
that no discrepancies are present between the CPU and GPU version of the
solver. As the linear solver, GMRES+LUSGS is used on the CPU, while GMRES with
AMG preconditioner and Diagonal Incomplete Lower Upper (DILU) smoother is used
for GPU-LA. Steady-state simulations are run for $10,000$ iterations. 
Convergence of integral quantities was monitored for the CPU and GPU-LA runs. In Fig. \ref{fig:crm_conv}, trends of drag and lift coefficients over the first 2500 iterations are shown. As can be noticed, the plots are almost superimposed, with very small differences shown in the upper and lower closeups on the right for the lift and drag coefficient, respectively. In both cases, a converged solution was already obtained after approximately 2000 iterations. This confirms that iteration after iteration, the solution is identical no matter which linear solver is used for linear algebra.


Then, the accuracy of the results is assessed considering both integral
quantities and pressure distributions across the aircraft surfaces.

\begin{figure}
  \includegraphics[width=.7\textwidth]{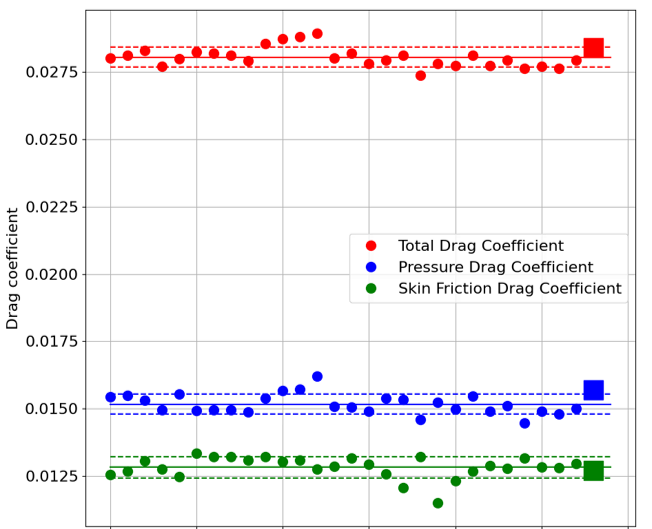}
  \caption{Predicted total, pressure and skin friction drag
    coefficients. Results of other solvers from DPW6 are denoted by the smaller
    dots, while OpenFOAM results are represented by the larger squares on the
    right of the figure.}
  \label{fig:crm_drag}
\end{figure}

Three versions of the drag coefficient are represented in figure
\ref{fig:crm_drag}, where total drag, pressure drag, and skin friction drag
are compared to the results of other workshop participants. Outliers for which
the predicted drag was too far from the average value (solid lines) are
excluded from the figure, so that the resultant standard deviation (dashed
lines) is quite small. The \textit{ICSFoam} findings, marked by squares, are
compared to all the other small dots representing the simulations executed
with other CFD software. In particular, the final values lie within the
standard deviation of the results from other participants and are considered
satisfactory by the authors. An angle of attack of $2.57$ degrees was used to
obtain the target lift coefficient, which is also in agreement with other
workshop participants. Pressure coefficients are extracted at six different
wing spanwise location an compared with available experimental data from NASA
Ames wind tunnel facility \cite{rivers2010}. We can notice that the agreement
is generally good throughout the wing, though some discrepancies are observed
near the shock location at $\eta=0.603$. This difference is completely
confirmed by the results from other participants \cite{tinoco2018}, where a
significant spread in the shock intensity and location has been observed,
which increases at higher angles of attack. These discrepancies are likely to
be due to a significant amount of shock buffeting, observed in the
experimental tests \cite{tinoco2018}, that cannot be captured by steady state
solutions.

\begin{figure}
  \centering \includegraphics[width=\textwidth]{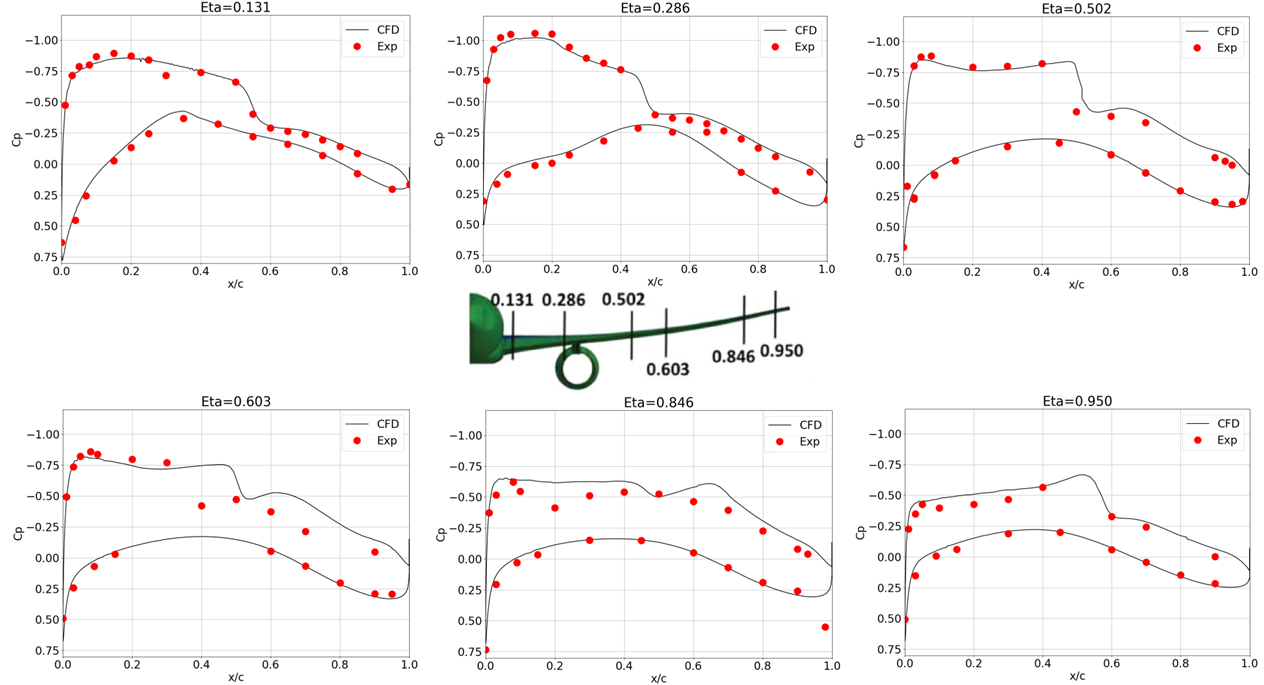}
  \caption{Comparison of experimental and computational pressure coefficient at
    different wing spans for NASA CRM test case.}
\end{figure}   

\subsubsection{Performance assessment}

Performance statistics are obtained as described in Sec.
\ref{sec:hardware}, and result from an average of measured computational times
over 500 iterations after the solution was converged. Simulations statistics
are shown in Table \ref{tab:absolute_times_mem_CRM} in terms of total wall
time and average on-chip memory memory usage per GPU. As already mentioned,
linear algebra is typically one of the most cumbersome parts of the iteration
in implicit solvers.

\begin{figure}
  \includegraphics[width=.8\textwidth]{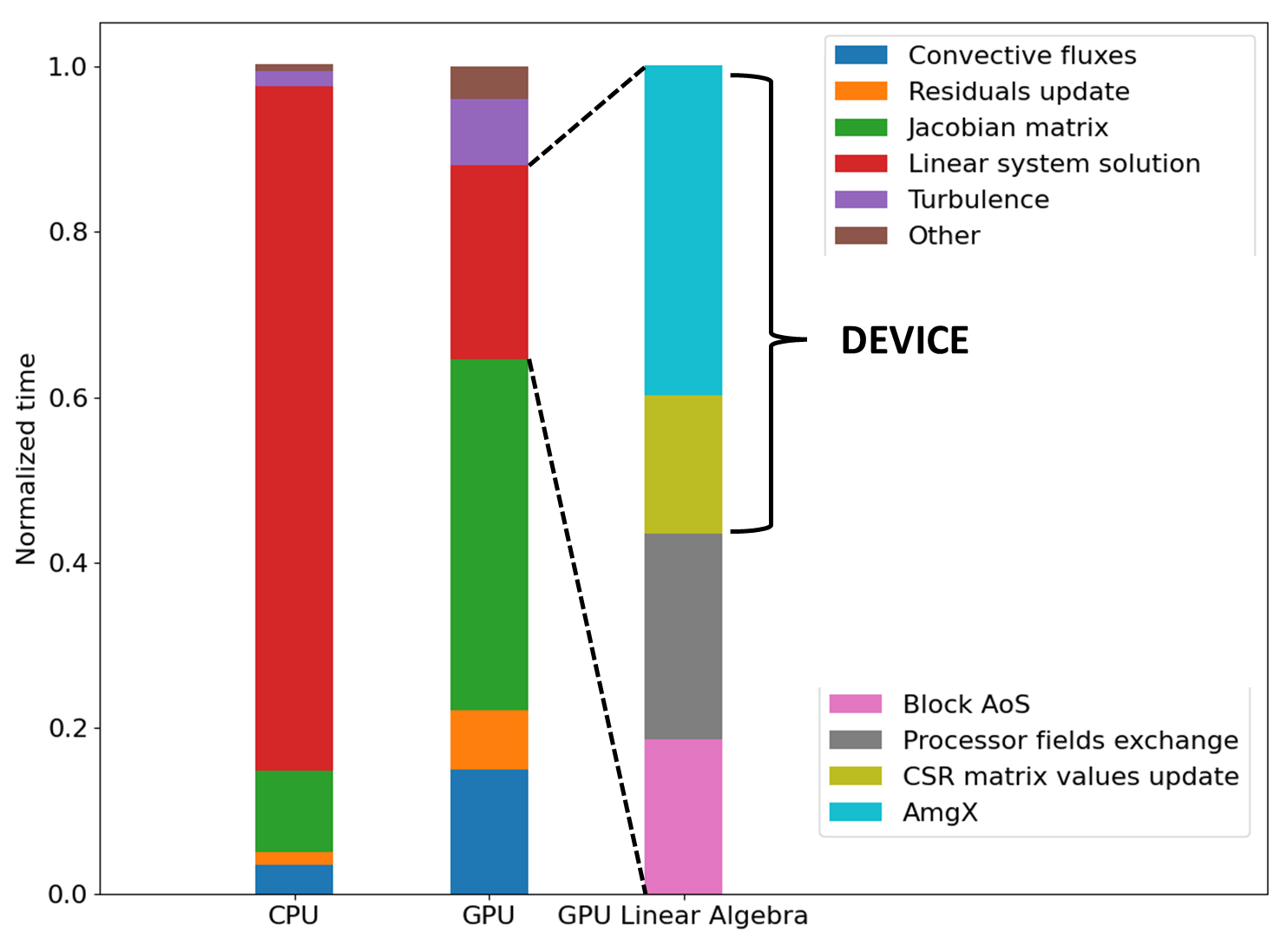}
  \caption{Decomposition of CPU and GPU iterations for the NASA CRM test
    case. Times are normalized with respective total iteration times. Last bar
    chart represents decomposition of the linear algebra step on the GPU.}
  \label{fig:CRM_iterDecomp}
\end{figure} 

For the density-based solver, this is testified by the first bar in
Fig. \ref{fig:CRM_iterDecomp}, showing the decomposition of a single iteration
into its main computational kernels for the CPU version of the code. This
figure reports the statistics obtained with the 260 million elements mesh
simulation on 18 nodes (864 CPU cores + 72 GPUs) using GMRES with AMG
preconditioner and DILU smoother as linear solver. We can see that almost $ 80
\%$ of the iteration is being spent solving the linear system. Although this
is somewhat a pessimistic perspective for general calculations, due to the
tight tolerance imposed on linear solver residuals (see Sec.
\ref{sec:hardware}), it can provide a good estimate for the initial phase of
the simulation. Indeed, when the simulation is far from convergence and high
Courant numbers are used (as typical of implicit solvers), the linear system
becomes very stiff and a large number of iterations are needed, significantly
impacting the total wall time. Instead, for the GPU-LA version, this is not
the case. Indeed, linear system solution now makes for about $25 \%$ of the
total iteration, while the most time consuming parts are related to the
Jacobian matrix assembly and convective fluxes calculation. Please notice that
for each bar, times are normalized with the corresponding total time relative
to that bar. For the GPU-LA solution, linear algebra can be further decomposed
in the preprocessing steps necessary to assemble and upload the matrix in the
CSR format needed by AmgX and the actual iterative solution. The latter and
the LDU to CSR conversion are performed on the device, while there is a
significant overhead due to MPI ranks communication and matrix conversion into
block AoS format. The figure clearly highlights a key point. First, future
work should concentrate on other computational kernels, as further
optimization of linear system solution would lead to a foreseeable performance
improvement in the range of $10-15 \%$. This would however require a much
higher implementation effort, with the additional risk of penalizing platform
portability and breaking compatibility with other software releases.

\begin{figure}
  \subfloat[][\emph{Overall}\label{fig:CRM_speedup_overall}]{\includegraphics[width=.47\textwidth]{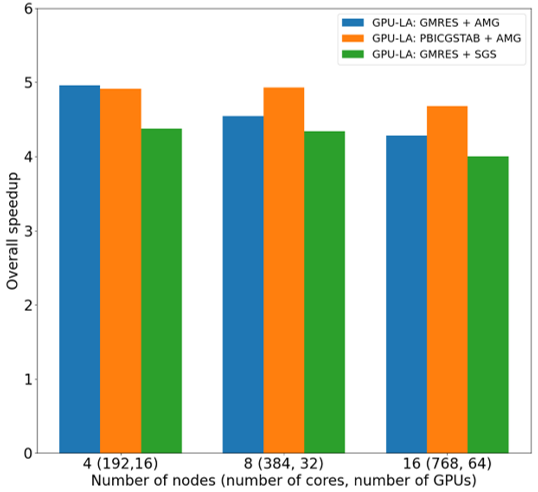}}
  \quad \subfloat[][\emph{Linear
      algebra}\label{fig:CRM_speedup_la}]{\includegraphics[width=.49\textwidth]{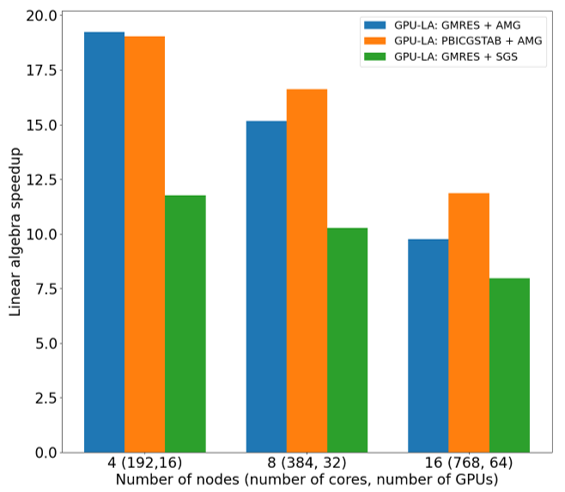}}
  \caption{Computational performance of GPU-LA in terms of speedup. Results are
    normalized with total iteration time (respectively, linear system solution
    time) for the CPU legacy case on the same number on nodes. }
  \label{fig:CRM_speedup}
\end{figure}  

Fig. \ref{fig:CRM_speedup} shows the obtained speedup for the 45 million
elements case and three different linear solvers, namely GMRES and PBiCGStab
with AMG preconditioner, and GMRES with SGS prconditioner. The latter
represents the configuration most similar to the one used in the CPU only
version. Fig. \ref{fig:CRM_speedup_overall} shows the overall speedup, while
Fig. \ref{fig:CRM_speedup_la} represents the speedup obtained for linear
algebra only, including the necessary preprocessing steps (red bar in
Fig. \ref{fig:CRM_iterDecomp}). Results are presented for 4, 8 and 16 GPU
nodes, and are normalized with total iteration time (respectively, linear
system solution time) for the CPU legacy case on the same number on nodes. The
overall simulation time is accelerated from 4 to 5 times, even as we approach
the strong scalability limit on 16 nodes. In addition, as expected, the usage
of multigrid preconditioner gives an edge compared to SGS, especially when the
number of cells per GPU is higher. The obtained performance improvement makes
the application of this method very appealing for coupled solvers.

\begin{figure}
  \includegraphics[width=.9\textwidth]{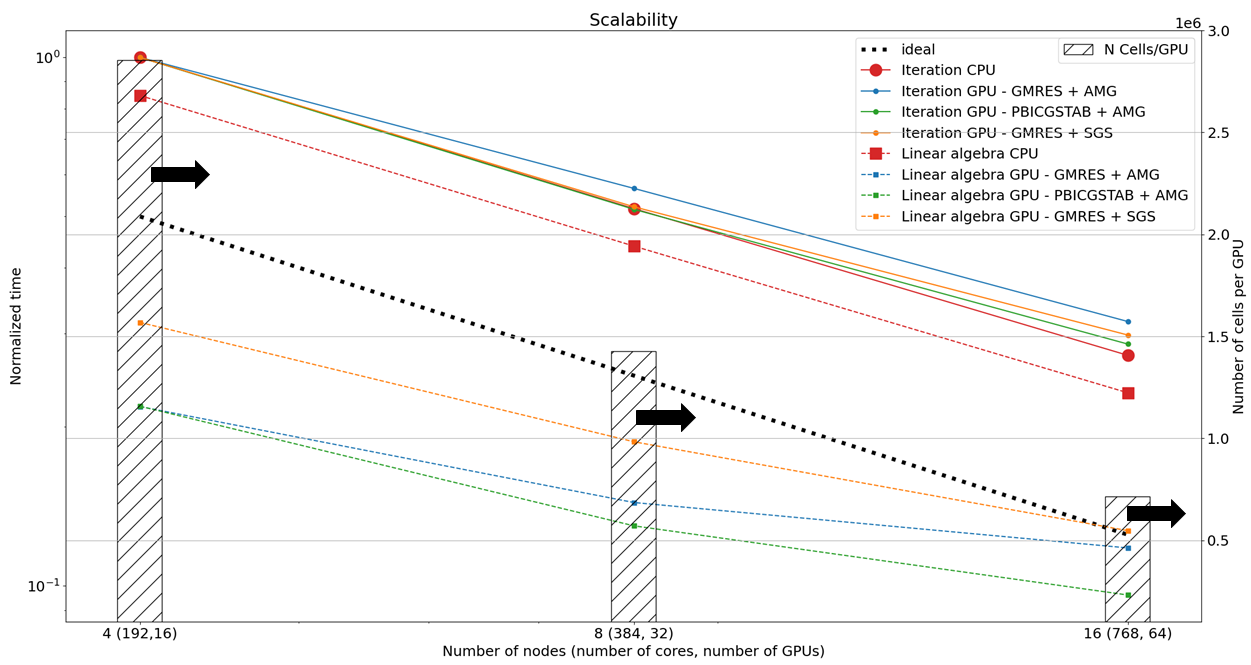}
  \caption{Computational performance of GPU-LA in terms of scalability. Results
    are normalized with total iteration time for CPU legacy case on 4 nodes (192
    cores).}
  \label{fig:CRM_scalability}
\end{figure}

Finally, Fig. \ref{fig:CRM_scalability} shows normalized iteration times on
different numbers of nodes, where the dotted black line shows the ideal
scalability trend. Reference is the total iteration time for the 4 nodes CPU
configuration. The bar charts represent the total number of cells per GPU. As
can be seen, PBiCGStab has the best performance in terms of scalability among
GPU solvers, showing a similar trend to the CPU solver, while for the four
nodes case GMRES has a slight edge. Since the number of streaming
multiprocessor in a single GPU is much higher than the number of CPU cores in
a node, it is to be expected that GPU-LA reaches a strong scalability limit
for a lower number of nodes than CPU only version. A number of cells around
1-5 million per GPU seems appropriate to exploit efficiently linear algebra
acceleration.

Table \ref{tab:absolute_times_mem_CRM} provides a quantitative overview of the
computational performance of the density based solver. The wall time spent
running 500 iterations is included for the CPU GMRES solver with LUSGS
preconditioner, and for both the GMRES and PBiCGStab GPU solvers with AMG
preconditioner. For the accelerated cases, the last column contains the
maximum VRAM usage divided by the number of GPUs.  These data can be employed
to compare the ICSFoam performance with proprietary and open-source CFD
software, demonstrating the general value of this research product. From an
industrial perspective, the stability of the coupled formulation is extremely
effective in relaxing the strict mesh requirements that come with standard
OpenFOAM solvers. Furthermore, the GPU acceleration provides a performance
boost that significantly lower the time-to-solution of applications typically
encountered in the aerospace sector, such as the NASA CRM presented in this
section.  An interesting observation can be extrapolated from the maximum VRAM
usage: the cases with 8 and 16 nodes are characterized by a low-to-medium GPU
memory load, considering a 40 GB limit for NVIDIA A100
accelerators. Consequently, future studies might regard the investigation of
different configurations to define the best compromise between hardware cost,
computational performance and number of nodes.

\begin{table}[]
  \centering
  \caption{Computational times for NASA CRM simulations}
  \label{tab:absolute_times_mem_CRM}
  \begin{tabular}{l|c|c|c|c}
    \multicolumn{1}{l|}{\textbf{Case}} & \multicolumn{1}{l|}{\textbf{Solver}} &
    \multicolumn{1}{l|}{\textbf{N GPUs}} & \multicolumn{1}{l|}{\textbf{WT x 500
        iter. {[}s{]}}} & \multicolumn{1}{p{0.2\textwidth}}{\textbf{Max VRAM /
        NGPUs {[}GB{]}}} \\ \hline \multicolumn{1}{l|}{} & \multicolumn{1}{l|}{} &
    \multicolumn{1}{l|}{} & \multicolumn{1}{l|}{} & \multicolumn{1}{l}{} \\ 45 M -
    4 nodes & CPU - GMRES + LUSGS & // & 15996 & // \\ & GPU - GMRES + AMG & 16 &
    3273 & 21.7 \\ & GPU - PBiCGStab + AMG & 16 & 3279 & 21.2
    \\ \multicolumn{1}{l|}{} & \multicolumn{1}{l|}{} & \multicolumn{1}{l|}{} &
    \multicolumn{1}{l|}{} & \multicolumn{1}{l}{} \\ 45 M - 8 nodes & CPU - GMRES +
    LUSGS & // & 8268 & // \\ & GPU - GMRES + AMG & 32 & 1830 & 13.9 \\ & GPU -
    PBiCGStab + AMG & 32 & 1699 & 13.7 \\ \multicolumn{1}{l|}{} &
    \multicolumn{1}{l|}{} & \multicolumn{1}{l|}{} & \multicolumn{1}{l|}{} &
    \multicolumn{1}{l}{} \\ 45 M - 16 nodes & CPU - GMRES + LUSGS & // & 4330 & //
    \\ & GPU - GMRES + AMG & 64 & 1045 & 9.9 \\ & GPU - PBiCGStab + AMG & 64 & 964
    & 9.7 \\ \multicolumn{1}{l|}{} & \multicolumn{1}{l|}{} & \multicolumn{1}{l|}{}
    & \multicolumn{1}{l|}{} & \multicolumn{1}{l}{} \\ 260 M - 18 nodes & GPU -
    GMRES +AMG & 72 & 7744 & 26.6
  \end{tabular}
\end{table}

\subsection{Test Case 2: incompressible DrivAer}

The second test case presented involves the stedy-state incompressible flow
field around the notchback version of the DrivAer from the series of
Automotive CFD Prediction Workshop \cite{autoCFD}. The workshop focuses on a
closed cooling configuration with static wheels and static floor. A
comprehensive set of experimental data from the Pininfarina Wind Tunnel
including aerodynamic forces and surface pressure measurements is available on
the website. The high-Reynolds mesh generated by the workshop committee is
used for the simulations. The domain is composed of a rectangular bounding box
of extension $\SI{120}{\meter} \times \SI{44}{\meter}$. The simulations are
"free air" and do not include the wind tunnel geometry. The mesh, generated
using ANSA BETA CAE, is composed of 128 million elements and has a $y^+\approx
30$. A fixed velocity value of $\SI{38.89}{\meter\per\second}$ is imposed at
the domain inlet, while a zero relative pressure is set at the outlet. Since
the mesh was originally generated for scale-resolving DES simulations, no
symmetry condition is imposed on the half plane and the full car span is
modeled. On the upper and side domain surfaces, a slip condition is
imposed. For RANS turbulence closure, the two-equations $k-\omega$ SST model
\cite{menter2003ten} is used. Second order upwind schemes are used for
convective fluxes.

Simulations were carried out with the coupled incompressible solver for 3000
iterations. Convergence was monitored by looking at integral aerodynamic
coefficients and residuals. Due to massive flow separation in the car wake, a
stationary solution could not be reached and flow statistics oscillates around
a mean value. For this reason, comparison of velocity contours with
experiments is performed using a velocity field averaged over the last 500
iterations. 

\begin{figure}
  \includegraphics[width=.9\textwidth]{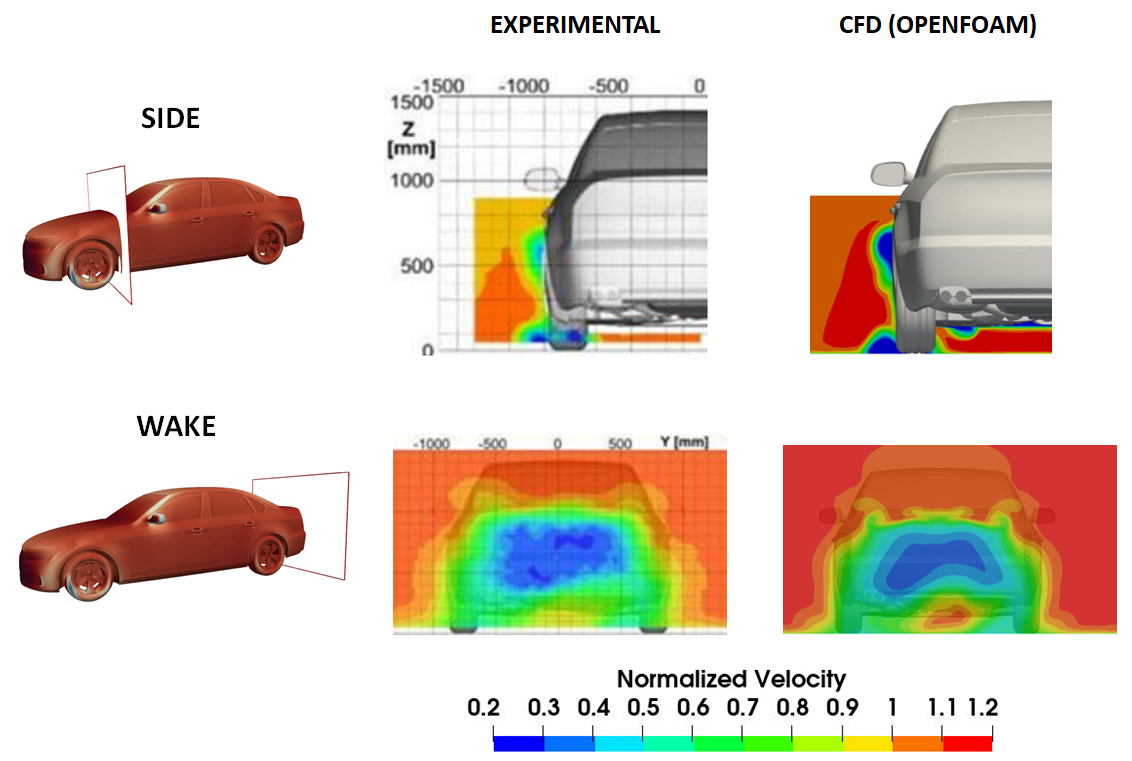}
  \caption{Normalized velocity contours at different planes for the DrivAer test
    case. Experimental results from Hupertz \etal \cite{hupertz2021} are
    compared with OpenFOAM.}
  \label{fig:drivaer_Uside}
\end{figure}

\begin{figure}
  \includegraphics[width=.8\textwidth]{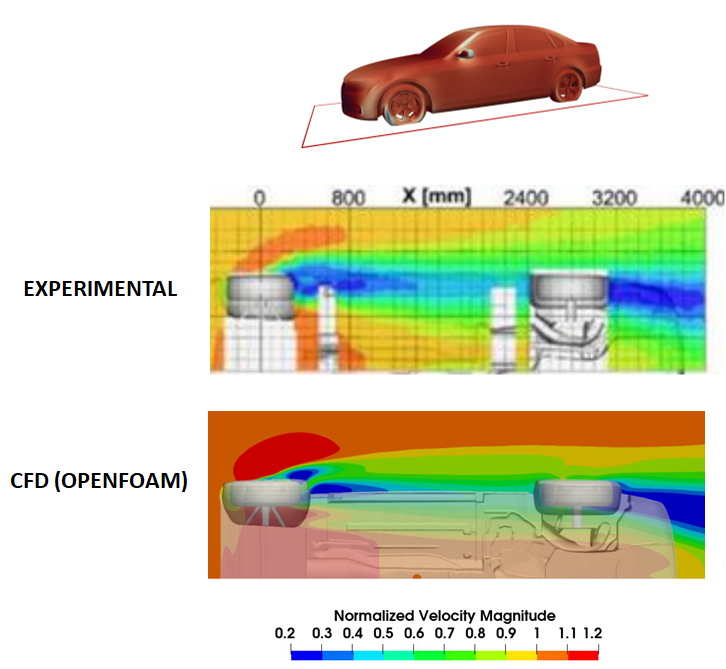}
  \caption{Normalized velocity contours at different planes for the DrivAer test
    case. Experimental results from Hupertz \etal \cite{hupertz2021} are
    compared with OpenFOAM.}
  \label{fig:drivaer_underbody}
\end{figure}

\begin{figure}
  \subfloat[][\emph{Underbody}]{\includegraphics[width=.47\textwidth]{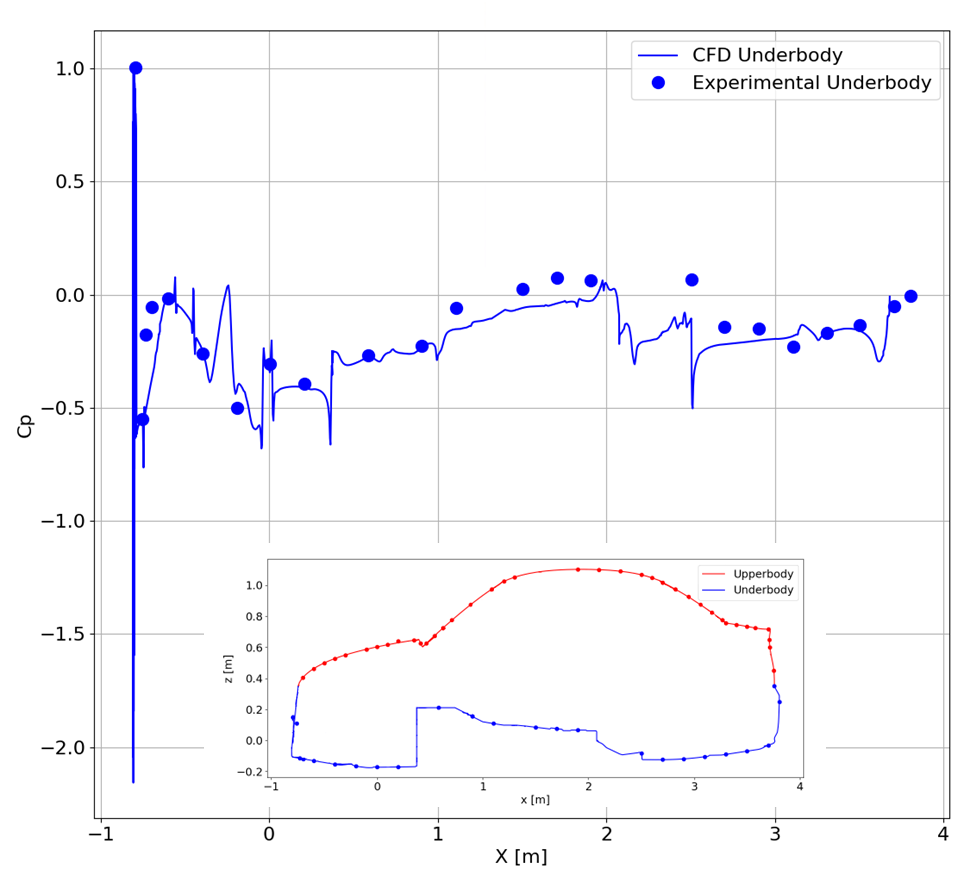}}
  \quad
  \subfloat[][\emph{Upperbody}]{\includegraphics[width=.48\textwidth]{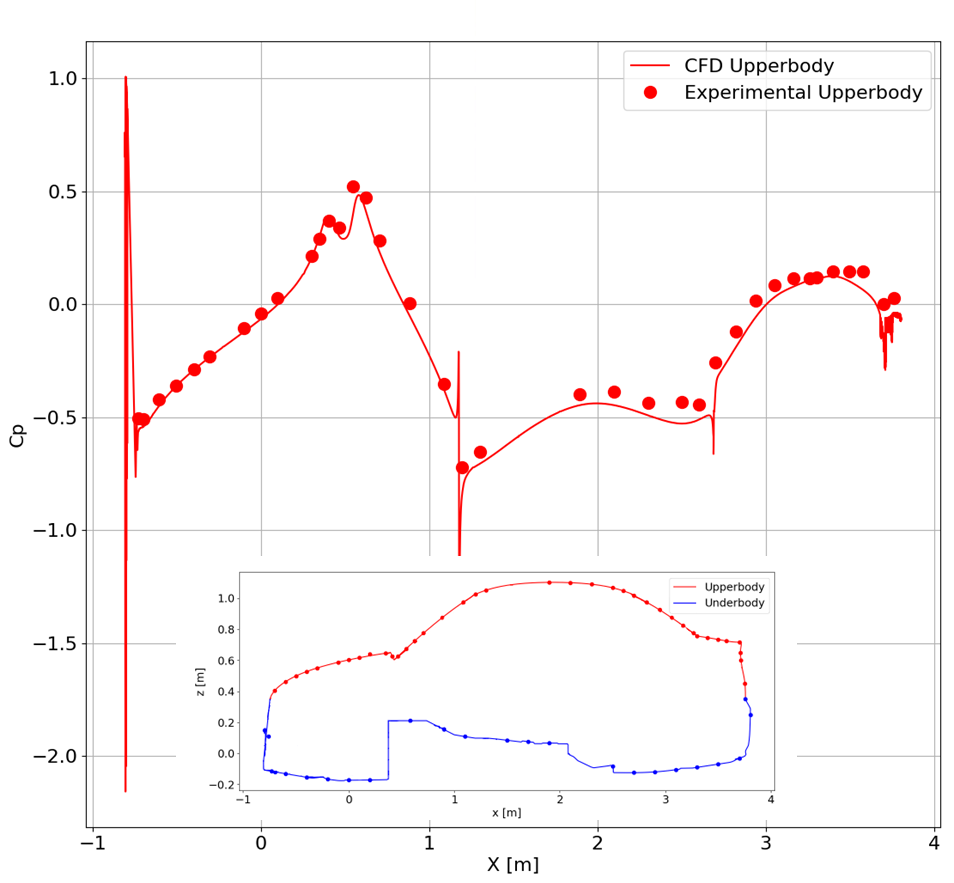}
  }
  \caption{Comparison of experimental and computational pressure coefficient on
    car symmetry plane at different probes location.}
  \label{fig:drivaer_cp}
\end{figure}

\begin{figure}
  \centering \subfloat[][\emph{Drag coefficient
      $C_D$}]{\includegraphics[width=.48\textwidth]{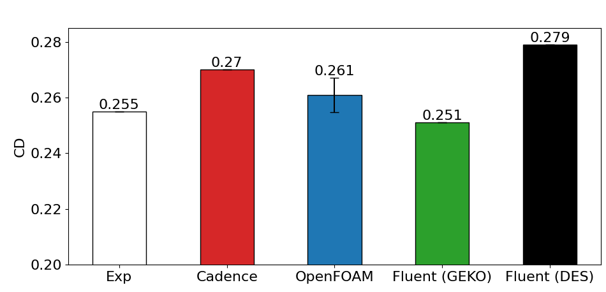}}
  \quad \subfloat[][\emph{Front lift coefficient
      $C_{L,f}$}]{\includegraphics[width=.48\textwidth]{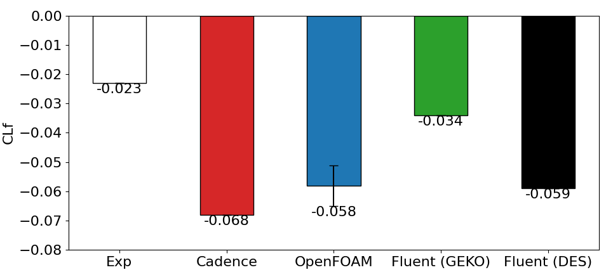}
  } \\ \subfloat[][\emph{Rear lift coefficient
      $C_{L,r}$}]{\includegraphics[width=.48\textwidth]{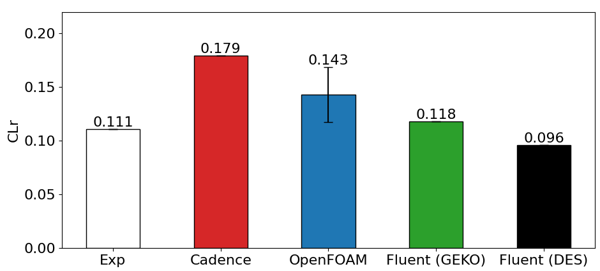}
  }
  \caption{Aerodynamic coefficients for the DrivAer case from 2nd Automotive CFD
    Workshop and OpenFOAM. Standard deviation of values over last 500 iterations
    is also reported on the OpenFOAM bar.}
  \label{fig:drivaer_coeffs}
\end{figure}

In Fig. \ref{fig:drivaer_Uside}, the comparison of mean velocity contours on the side and in the main wake area is reported. A satisfactory agreement is found between CFD and experimental
findings, including the side wake, wheels wake close to the ground, as well as
the underbody wake in the rear part of the vehicle. The main features of the
underbody flow are also correctly captures, as testified by
Fig. \ref{fig:drivaer_underbody}. The extension of wheel wakes is comparable,
although the front wheel wake is slightly deflected outboard. Experimental
measurements from pressure probes on the symmetry plane are also available for
the upperbody and the underbody of the vehicle. Probes locations are shown on
the bottom of the plots in Fig. \ref{fig:drivaer_cp}. As can be seen from the
figure, there is a very good agreement with experimental values on both sides
of the car, especially on the hood, where the flow is mostly attached to the
wall. Pressure coefficient is slightly underpredicted on the roof, above the
trunk and in the separation area in the underbody. Finally, in
Fig. \ref{fig:drivaer_coeffs}, drag, front and rear lift coefficients are
compared with the ones obtained with commercial CFD solvers by other
participants to the workshop. Standard deviation of values over last 500
iterations is also reported on the OpenFOAM bar. As can be noticed, there is a
significant dispersion in the values obtained with CFD solvers. Drag values
are within the $10 \%$ of experimental range, while a general disagreement
between CFD lift coefficients and experimental data is obtained, especially
for what concerns the front lift.


\subsubsection{Performance assessment}

\begin{table}[]
  \centering
  \caption{Computational times for DrivAer simulations}
  \label{tab:absolute_times_mem_drivaer}
  \begin{tabular}{l|c|c|c|c}
    \multicolumn{1}{l|}{\textbf{Case}} & \multicolumn{1}{l|}{\textbf{Solver}} &
    \multicolumn{1}{l|}{\textbf{N GPUs}} & \multicolumn{1}{l|}{\textbf{WT x 500
        iter. {[}s{]}}} & \multicolumn{1}{p{0.2\textwidth}}{\textbf{Max VRAM /
        NGPUs {[}GB{]}}} \\ \hline \multicolumn{1}{l|}{} & \multicolumn{1}{l|}{} &
    \multicolumn{1}{l|}{} & \multicolumn{1}{l|}{} & \multicolumn{1}{l}{} \\ 128 M
    - 8 nodes & CPU - Segregated & // & 2389 & // \\ & GPU - GMRES + AMG & 32 &
    2395 & 24.35 \\ & GPU - PBiCGStab + AMG & 32 & 2434 & 23.07
    \\ \multicolumn{1}{l|}{} & \multicolumn{1}{l|}{} & \multicolumn{1}{l|}{} &
    \multicolumn{1}{l|}{} & \multicolumn{1}{l}{} \\ 128 M - 16 nodes & CPU -
    Segregated & // & 1378 & // \\ & GPU - GMRES + AMG & 64 & 1383 & 15.17 \\ &
    GPU - PBiCGStab + AMG & 64 & 1253 & 14.73
  \end{tabular}
\end{table}

This section includes a performance comparison between the coupled solver and
the segregated pressure-based solver \textit{simpleFoam} released with
OpenFOAM vanilla. Simulations statistics are shown in Table
\ref{tab:absolute_times_mem_drivaer} in terms of total wall time and average
on-chip memory memory usage per GPU. Generally, the memory requirement is
lower than the one of NASA CRM, due to the absence of the energy equation in
the coupled system.

\begin{figure}
  \includegraphics[width=.8\textwidth]{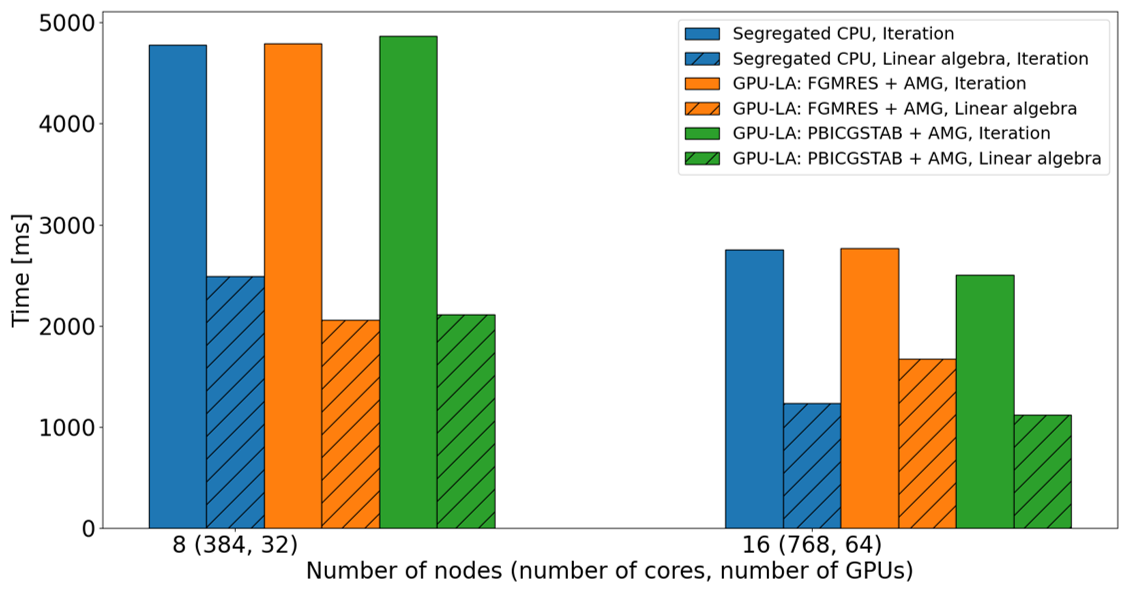}
  \caption{Computational times for the Drivaer test case. Average times spent
    for the whole iteration as well as linear algebra only are shown for
    segregated CPU and GPU-LA configurations.}
  \label{fig:drivaer_compTimes}
\end{figure}

It must be remarked that convergence could not be obtained with
\textit{simpleFoam} due to stability issues, even employing first order
schemes and very low relaxation factors. On the contrary, the coupled solver
was characterized by a smooth convergence history, highlighting the strong
benefits of a full pressure-velocity coupling. Therefore, to make comparisons,
the simulation was run until convergence with the coupled solver and then for
other 500 iterations with the SIMPLE approach to gather the statistics. For
\textit{simpleFoam}, the native Geometric Algebraic MultiGrid (GAMG) is used
for the pressure equation, while PBiCGStab with DILU preconditioner is used
for the momentum equation. Due to mesh non-orthogonality, the pressure
correction equation is typically solved multiple times in the SIMPLE approach,
so that explicit non orthogonal corrections can be included in the
pressure-velocity coupling. In the present study, one additional
non-orthogonal correction was employed for the pressure equation. It is
worthwhile to notice that segregated solvers usually present a significantly
lower time per single iteration compared to the coupled
approach. Fig. \ref{fig:drivaer_compTimes} shows that with GPU acceleration of
linear algebra, the authors benefit from the stability of a fully implicit
approach, still maintaining a time per iteration comparable, if not lower, to
those of the segregated solver. As for the CRM simulation, the PBiCGStab
solver shows better scalability compared to GMRES.

\begin{figure}
  \includegraphics[width=.8\textwidth]{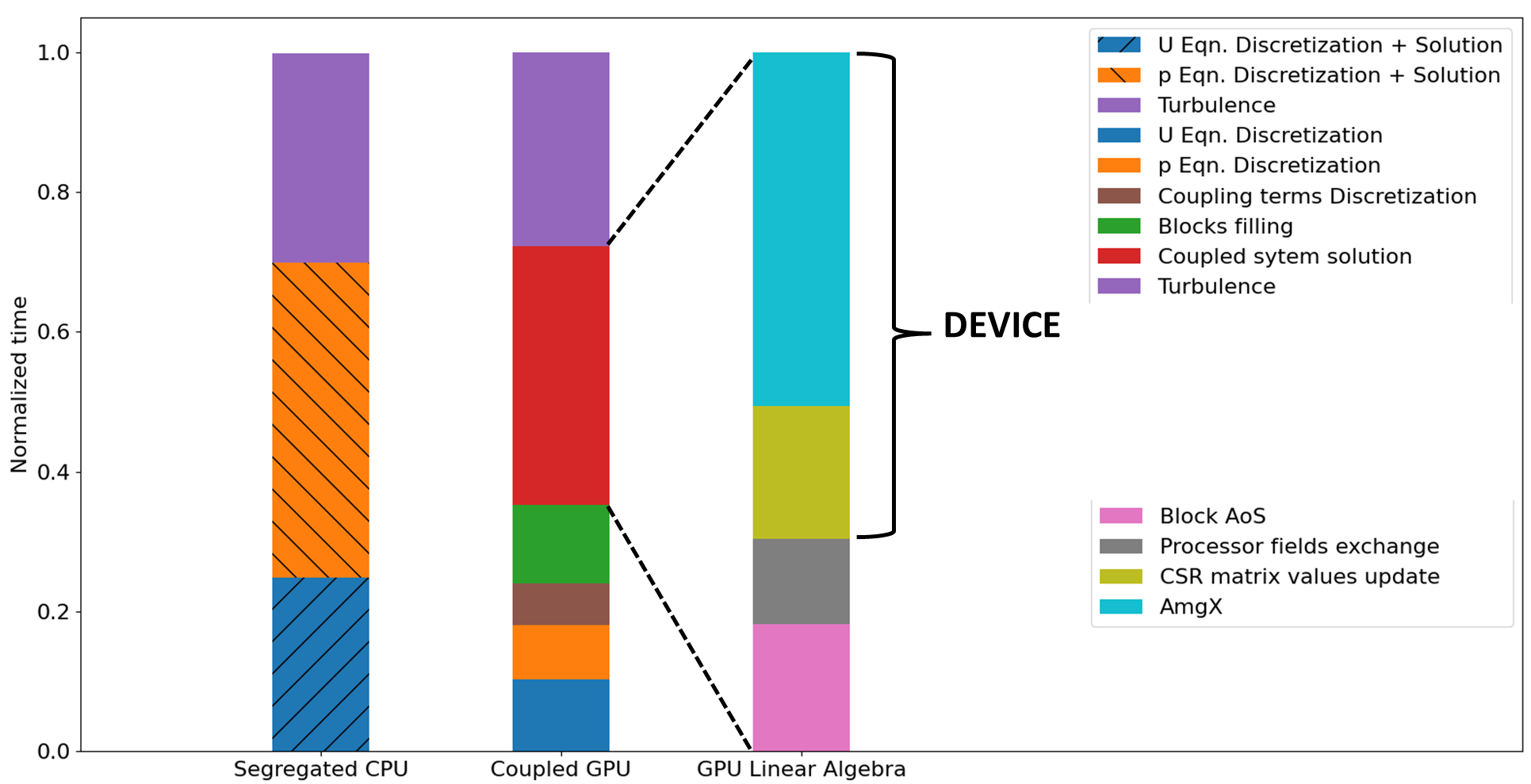}
  \caption{Decomposition of CPU and GPU iterations for the DrivAer test
    case. Times are normalized with respective total iteration times.  Last bar
    chart represents decomposition of the linear algebra step on the GPU.}
  \label{fig:drivaer_iterDecomp}
\end{figure}

Fig. \ref{fig:drivaer_iterDecomp} illustrates the decomposition of the
iteration for the two solvers into the main computational kernels. Times are
again normalized with respect to the total iteration times. For
\textit{simpleFoam}, the iteration is simply composed of momentum predictor,
pressure correction equation and turbulence solution. Pressure equation
discretization and solution times in the first bar include also the
non-orthogonal correction. For the coupled solver, about one third of the
iteration is used to discretize and assemble the coupled matrix. The
additional coupling terms need to be discretized as described in Sec.
\ref{sec:met} and inserted in the matrix, leading to an overhead. This,
however, is more than compensated by the faster linear algebra solution
obtained through the offloading onto the device. With a two-equation
turbulence model, the solution of k and $\omega$ represents approximately $25
\%$ of total iteration time.

\begin{figure}
\centering
  \includegraphics[width=.8\textwidth]{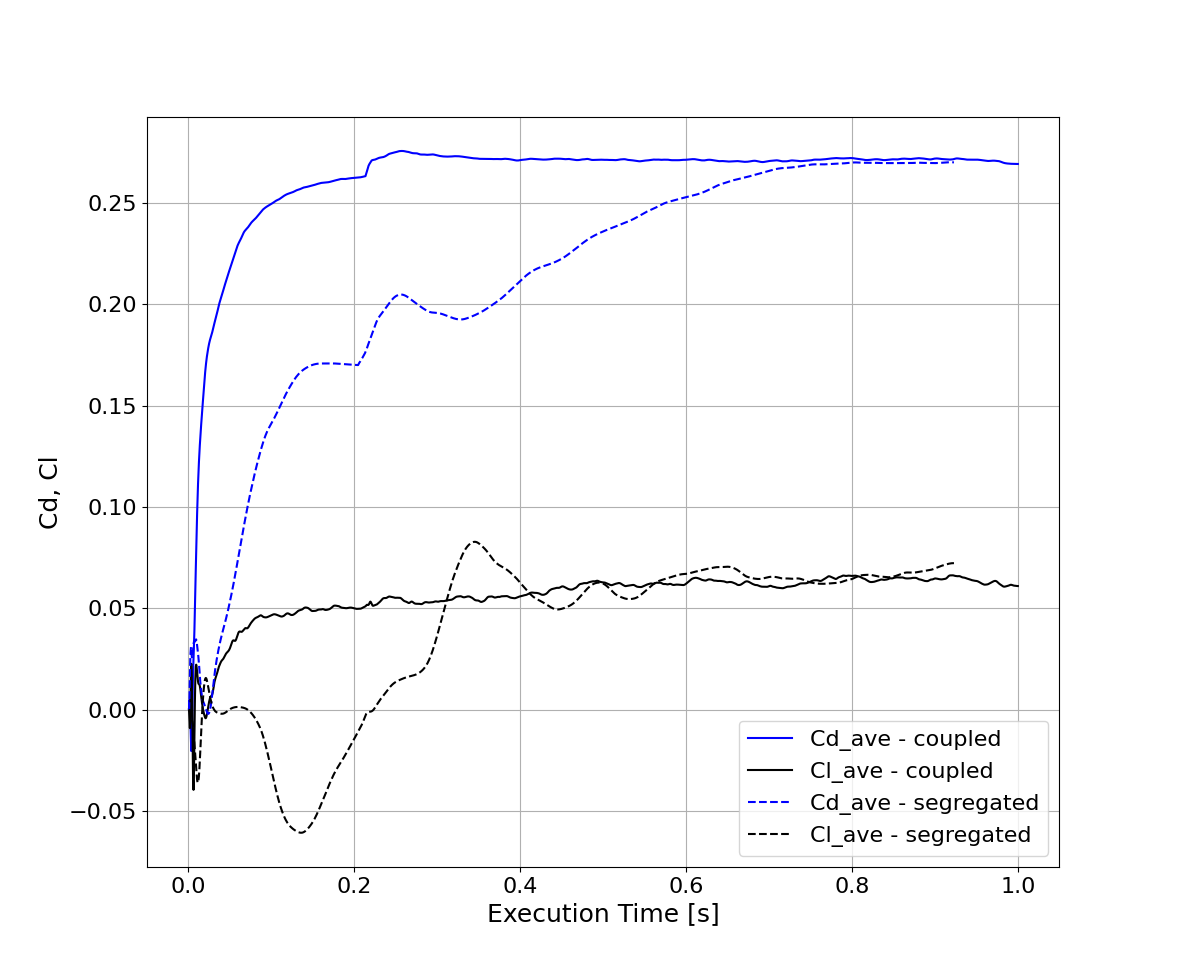}
  \caption{Convergence of aerodynamic coefficients for the 22 M elements test. Coupled vs segregated solvers in terms of total simulation time.}
  \label{fig:drivaer_conv_coupl_segr}
\end{figure} 

Since convergence could not be obtained with the segregated solver, an additional test was performed to demonstrate the benefit of using a coupled solver with linear algebra offloading in terms of overall simulation performance. A coarser mesh of 22 million elements was generated using \textit{snappyHexMesh}, the built-in OpenFOAM mesher. This tool is still far from being sustainable in industrial workflows, but allows to obtain a mesh with an improved quality according to OpenFOAM standards. Standard simulation settings were adopted, as previously described. Calculations were performed using a single GPU node, with 48 CPU cores and 4 GPUs (for coupled solver only). Fig. \ref{fig:drivaer_conv_coupl_segr} shows the convergence of averaged aerodynamic coefficients versus total simulation times for the two solvers. A window average over 400 iterations was employed in this case to assess the convergence of the solution. It can be seen that the coupled solver reaches a converged value of the average drag coefficient approximately three times faster than the segregated solver in terms of total wall time. Average lift coefficient also shows an improved convergence trend.

\section{Conclusions}
\label{sec:concl}

This paper has presented a workflow to extend a generic library for implicit
coupled simulations through GPU acceleration of linear algebra. The
OpenFOAM-based \textit{ICSFoam} library for finite-volume discretization of
coupled systems of PDEs, was coupled with NVIDIA AmgX through a modified
version of the \textit{amgx4Foam} library. This allowed to offload the most
time consuming part of the iteration onto the accelerator while keeping the
remaining part on the host, in a heterogeneous CPU/GPU architecture
paradigm. We showed that significant performance speedups can be easily
achieved. In particular, speedups of the order of $10-20x$ for the linear
algebra only and $4-5x$ overall have been obtained. The offloading of sparse
matrices for coupled linear systems gives in general better performance with
respect to segregated solvers due to the more favourable arithmetic intensity
of the block AoS structure, while retaining a low level of implementation
effort. This fact, combined with the improved convergence and the inherent
robustness and stability of coupled solvers, shows that this approach is
suitable for industrial simulations of complex geometries. To prove this fact,
we show the results for a compressible and incompressible test cases using
pressure and density-based solvers, respectively.\\ Remarkably, this is the
first implementation of GPU accelerated coupled implicit simulations in
OpenFOAM, which is arguably the most widely employed open source CFD software
in academic and industrial context. We highlighted that, with the current
implementation, linear algebra of the coupled solver is not the most
time-consuming part of the iteration anymore, weighting now for approximately
$20-25 \%$ of total wall time.\\ Therefore, future efforts to further
accelerate the code need necessarily to focus on the offloading of the other
computational kernels, such as discretization, fluxes calculation and
turbulence equations. We also showed that, despite future generation GPU
devices will feature more and more on-chip memory, VRAM can be still a
critical issue for large-sized industrial problems. This can severely impact
the minimum number of GPUs required for a single simulation, with consequent
increase in the hardware cost. Optimization needs to be carried out to reduce
the memory footprint of linear algebra algorithms. \\ Finally, we can expect
even superior performances when CPU-GPU shared-memory architectures
(e.g. NVIDIA Grace Hopper superchip) will be available, since memory copies
between host and devices systems are no longer needed

\backmatter








\section*{Declarations}

\subsection*{Conflict of interest}
The authors declare no conflict of interest.
\subsection*{Availability of data and materials}
The data presented in this study are available upon reasonable request from
the corresponding author.

\bibliography{ref}

\begin{thebibliography}{10}
\providecommand{\url}[1]{{#1}}
\providecommand{\urlprefix}{URL }
\providecommand{\doi}[1]{\url{https://doi.org/#1}}
\bibcommenthead

\bibitem{rivers2010}
M.B. Rivers, A.~Dittberner, Experimental investigations of the nasa common
  research model.
\newblock Journal of Aircraft \textbf{51}(4), 1183--1193 (2014).
\newblock \doi{10.2514/1.C032626}

\bibitem{2014-01-0590}
D.~Wood, M.A. Passmore, A.K. Perry, {Experimental Data for the Validation of
  Numerical Methods - SAE Reference Notchback Model}.
\newblock {SAE International Journal of Passenger Cars - Mechanical Systems}
  \textbf{7}(1), 145--154 (2014).
\newblock \doi{https://doi.org/10.4271/2014-01-0590}

\bibitem{6270749}
J.M. Cebri'n, G.D. Guerrero, J.M. García, in \emph{2012 IEEE 26th
  International Parallel and Distributed Processing Symposium Workshops \& PhD
  Forum} (2012), pp. 1014--1022.
\newblock \doi{10.1109/IPDPSW.2012.124}

\bibitem{Price2016}
D.~Price, M.~Clark, e.a. Barsdell, B.R., {Optimizing performance-per-watt on
  GPUs in high performance computing}.
\newblock Comput Sci Res \textbf{31}, 185--193 (2016).
\newblock \doi{https://doi.org/10.1007/s00450-015-0300-5}

\bibitem{KRZYWANIAK2023396}
A.~Krzywaniak, P.~Czarnul, J.~Proficz, Dynamic gpu power capping with online
  performance tracing for energy efficient gpu computing using depo tool.
\newblock Future Generation Computer Systems \textbf{145}, 396--414 (2023).
\newblock \doi{https://doi.org/10.1016/j.future.2023.03.041}

\bibitem{Ansys}
Ansys.
\newblock {Fluent Fluid Simulation Software} (2023).
\newblock \url{https://www.ansys.com/products/fluids/ansys-fluent} [Accessed:
  Dec 2023]

\bibitem{starCCM}
SIEMENS.
\newblock {Simcenter STAR-CCM+ CFD software} (2023).
\newblock
  \url{https://plm.sw.siemens.com/en-US/simcenter/fluids-thermal-simulation/star-ccm/}
  [Accessed: Dec 2023]

\bibitem{naumov2015}
M.~Naumov, M.~Arsaev, P.~Castonguay, J.~Cohen, J.~Demouth, J.~Eaton, S.~Layton,
  N.~Markovskiy, I.~Reguly, N.~Sakharnykh, V.~Sellappan, R.~Strzodka, Amgx: A
  library for gpu accelerated algebraic multigrid and preconditioned iterative
  methods.
\newblock SIAM Journal on Scientific Computing \textbf{37}(5), S602--S626
  (2015).
\newblock \doi{10.1137/140980260}

\bibitem{piscaglia2023}
F.~Piscaglia, F.~Ghioldi, {GPU Acceleration of CFD Simulations in OpenFOAM}.
\newblock Aerospace \textbf{10}(9) (2023).
\newblock \doi{10.3390/aerospace10090792}

\bibitem{WILLIAMS2016}
J.~Williams, C.~Sarofeen, H.~Shan, M.~Conley, An accelerated iterative linear
  solver with gpus for cfd calculations of unstructured grids.
\newblock Procedia Computer Science \textbf{80}, 1291--1300 (2016).
\newblock \doi{https://doi.org/10.1016/j.procs.2016.05.504}.
\newblock International Conference on Computational Science 2016, ICCS 2016,
  6-8 June 2016, San Diego, California, USA

\bibitem{ZHANG2023}
X.~Zhang, X.~Guo, Y.~Weng, X.~Zhang, Y.~Lu, Z.~Zhao, Hybrid mpi and cuda
  paralleled finite volume unstructured cfd simulations on a multi-gpu system.
\newblock Future Generation Computer Systems \textbf{139}, 1--16 (2023).
\newblock \doi{https://doi.org/10.1016/j.future.2022.09.005}

\bibitem{jaiswal2016}
S.~Jaiswal, R.~Reddy, R.~Banerjee, S.~Sato, D.~Komagata, M.~Ando, J.~Okada, in
  \emph{2016 IEEE 23rd International Conference on High Performance Computing
  Workshops (HiPCW)} (2016), pp. 81--89.
\newblock \doi{10.1109/HiPCW.2016.020}

\bibitem{Peric}
J.H. Ferziger, M.~Peri\'{c}, R.L. Street, \emph{{Computational Methods for
  Fluid Dynamics}}, 4th edn. (Springer, 2020)

\bibitem{mangani2014}
M.B. L.~Mangani, M.~Darwish, Development of a novel fully coupled solver in
  openfoam: Steady-state incompressible turbulent flows.
\newblock Numerical Heat Transfer, Part B: Fundamentals \textbf{66}(1), 1--20
  (2014).
\newblock \doi{10.1080/10407790.2014.894448}

\bibitem{mangani2016}
M.D. L.~Mangani, F.~Moukalled, An openfoam pressure-based coupled cfd solver
  for turbulent and compressible flows in turbomachinery applications.
\newblock Numerical Heat Transfer, Part B: Fundamentals \textbf{69}(5),
  413--431 (2016).
\newblock \doi{10.1080/10407790.2015.1125212}

\bibitem{DARWISH2018}
M.~Darwish, L.~Mangani, F.~Moukalled, Implicit boundary conditions for coupled
  solvers.
\newblock Computers \& Fluids \textbf{168}, 54--66 (2018)

\bibitem{nvidiaH100}
\url{"https://www.nvidia.com/it-it/data-center/h100"}.
\newblock NVIDIA GPU H100 Tensor Core

\bibitem{weller1998}
H.G. Weller, G.~Tabor, H.~Jasak, C.~Fureby, A tensorial approach to
  computational continuum mechanics using object-oriented techniques.
\newblock Computers in Physics \textbf{12}(6), 620--631 (1998)

\bibitem{audiOF}
\url{"https://www.esi-group.com/customer-successes/audi-maximizes..."}.
\newblock Audi Maximizes the Driving Range and Acoustic Quality of the New
  e-tron Using ESI Virtual Prototyping Solutions

\bibitem{OLIANI2023}
S.~Oliani, N.~Casari, M.~Carnevale, Icsfoam: An openfoam library for implicit
  coupled simulations of high-speed flows.
\newblock Computer Physics Communications \textbf{286}, 108673 (2023).
\newblock \doi{10.1016/j.cpc.2023.108673}

\bibitem{UROIC2021}
T.~Uroić, H.~Jasak, Parallelisation of selective algebraic multigrid for
  block–pressure–velocity system in openfoam.
\newblock Computer Physics Communications \textbf{258}, 107529 (2021).
\newblock \doi{https://doi.org/10.1016/j.cpc.2020.107529}

\bibitem{DARWISH2009}
M.~Darwish, I.~Sraj, F.~Moukalled, A coupled finite volume solver for the
  solution of incompressible flows on unstructured grids.
\newblock Journal of Computational Physics \textbf{228}(1), 180--201 (2009).
\newblock \doi{https://doi.org/10.1016/j.jcp.2008.08.027}

\bibitem{OpenFOAM-com}
{ESI OpenCFD OpenFOAM}.
\newblock \url{http://www.openfoam.com/}

\bibitem{OpenFOAM-dev}
{The OpenFOAM Foundation}.
\newblock \url{http://www.openfoam.org/dev.php}

\bibitem{GHIOLDI2023}
F.~Ghioldi, F.~Piscaglia, {Acceleration of supersonic/hypersonic reactive CFD
  simulations via heterogeneous CPU-GPU supercomputing}.
\newblock Computers \& Fluids \textbf{266}, 106041 (2023).
\newblock \doi{https://doi.org/10.1016/j.compfluid.2023.106041}

\bibitem{heyns2014}
J.~Heyns, O.~Oxtoby, Modelling high-speed viscous flow in openfoam®.
\newblock 9th South African Conference on Computational and Applied Mechanics,
  SACAM 2014  (2014)

\bibitem{oliani2022HBM}
S.~Oliani, N.~Casari, M.~Carnevale, A new framework for the harmonic balance
  method in openfoam.
\newblock Machines \textbf{10}(4) (2022)

\bibitem{fadiga2023}
E.~Fadiga, F.~Rondina, S.~Oliani, T.~Benacchio, D.~Malacrida, L.~Capone,
  (2023).
\newblock \doi{10.23967/c.coupled.2023.018}

\bibitem{saad1986}
Y.~Saad, M.H. Schultz, Gmres: a generalized minimal residual algorithm for
  solving nonsymmetric linear systems.
\newblock Siam Journal on Scientific and Statistical Computing \textbf{7},
  856--869 (1986)

\bibitem{jameson1981}
A.~Jameson, W.~Schmidt, E.~Turkel, Solutions of the euler equations by finite
  volume methods using runge-kutta time-stepping schemes.
\newblock AIAA paper \textbf{1259} (1981)

\bibitem{luo1998}
H.~Luo, J.D. Baum, R.~Löhner, A fast, matrix-free implicit method for
  compressible flows on unstructured grids.
\newblock Journal of Computational Physics \textbf{146}(2), 664--690 (1998)

\bibitem{blazek2015}
J.~Blazek, \emph{Computational Fluid Dynamics: Principles and Applications},
  3rd edn. (Butterworth-Heinemann, 2015)

\bibitem{patankar1980}
S.~Patankar, \emph{Numerical Heat Transfer and Fluid Flow}, 1st edn. (Taylor
  and Francis Group, 1980)

\bibitem{ISSA1986}
R.~Issa, Solution of the implicitly discretised fluid flow equations by
  operator-splitting.
\newblock Journal of Computational Physics \textbf{62}(1), 40--65 (1986)

\bibitem{rhie1983}
C.M. Rhie, W.L. Chow, Numerical study of the turbulent flow past an airfoil
  with trailing edge separation.
\newblock AIAA Journal \textbf{21}(11), 1525--1532 (1983).
\newblock \doi{10.2514/3.8284}

\bibitem{martineau2020}
M.~Martineau, S.~Posey, F.~Spiga, in \emph{8th OpenFOAM Conference} (2020)

\bibitem{davinci-1}
\url{"https://www.leonardo.com/en/innovation-technology/davinci-1"}.
\newblock Supercomputer davinci-1

\bibitem{vassberg2008}
J.~Vassberg, M.~Dehaan, M.~Rivers, R.~Wahls, in \emph{26th AIAA Applied
  Aerodynamics Conference} (2008).
\newblock \doi{10.2514/6.2008-6919}

\bibitem{tinoco2018}
E.N. Tinoco, O.P. Brodersen, S.~Keye, K.R. Laflin, E.~Feltrop, J.C. Vassberg,
  M.~Mani, B.~Rider, R.A. Wahls, J.H. Morrison, D.~Hue, C.J. Roy, D.J.
  Mavriplis, M.~Murayama, Summary data from the sixth aiaa cfd drag prediction
  workshop: Crm cases.
\newblock Journal of Aircraft \textbf{55}(4), 1352--1379 (2018).
\newblock \doi{10.2514/1.C034409}

\bibitem{dpw6}
\url{"https://aiaa-dpw.larc.nasa.gov/Workshop6/workshop6.html"}.
\newblock 6th AIAA CFD Drag Prediction Workshop

\bibitem{spalart1992}
P.~Spalart, S.~Allmaras, \emph{A one-equation turbulence model for aerodynamic
  flows} (1992).
\newblock \doi{10.2514/6.1992-439}

\bibitem{autoCFD}
\url{https://autocfd.eng.ox.ac.uk/}.
\newblock Automotive CFD Prediction Workshop

\bibitem{menter2003ten}
F.R. Menter, M.~Kuntz, R.~Langtry, Ten years of industrial experience with the
  sst turbulence model.
\newblock Turbulence, heat and mass transfer \textbf{4}(1), 625--632 (2003)

\bibitem{hupertz2021}
B.~Hupertz, K.~Chalupa, L.~Krueger, K.~Howard, H.D. Glueck, N.~Lewington, J.H.
  Chang, Y.s. Shin,  (2021).
\newblock \doi{10.4271/2021-01-0958}

\end{thebibliography}

\end{document}